\documentclass[11pt, a4paper]{article}
\usepackage{jheppub}
\usepackage{amsmath}

\newcommand{\cO}{{\cal O}}

\usepackage{graphicx}
\usepackage{amsmath}
\usepackage{amsfonts}
\usepackage{epsfig}
\usepackage{slashed}
\usepackage{braket}
\usepackage{multirow}
\usepackage{comment}
\usepackage{cases}
\usepackage{color}
\usepackage{bookmark}

\usepackage{tikz}
\usepackage{smartdiagram}
\usetikzlibrary{decorations.pathmorphing}
\usesmartdiagramlibrary{additions} 
\usetikzlibrary{arrows,arrows.meta,graphs} 
\usetikzlibrary{positioning,shapes.misc,shadows}
\usepackage{tabularray}
\NewTblrEnviron{spectblr}
\SetTblrOuter[spectblr]{long}
\SetTblrInner[spectblr]{
	hlines = {fg=white,wd=2pt}, colsep = 2pt,
	row{1} = {fg=white, bg=azure3, font=\bfseries\sffamily},
	rowhead = 1,
}
\allowdisplaybreaks

\def\nn{\nonumber}

\newcommand{\cA}{{\cal A}}

\newcommand{\cS}{{\cal S}}

\newcommand{\GS}{\mathcal G_{\mathcal S}}
\newcommand{\FS}{\mathcal F_{\mathcal S}}
\newcommand{\GT}{\mathcal G_{\mathcal T}}
\newcommand{\FT}{\mathcal F_{\mathcal T}}
\newcommand{\bGS}{\bar{\mathcal G}_{\mathcal S}}
\newcommand{\bFS}{\bar{\mathcal F}_{\mathcal S}}
\newcommand{\bGT}{\bar{\mathcal G}_{\mathcal T}}
\newcommand{\bFT}{\bar{\mathcal F}_{\mathcal T}}

\newcommand{\btTh}{\bar{\Theta}}
\newcommand{\btSig}{\bar{\Sigma}}
\newcommand{\bPsi}{\bar{\Psi}}
\newcommand{\bg}{\bar{g}}
\newcommand{\bcD}{\bar{\mathcal D}}
\newcommand{\dbchik}{\delta\bar{\chi}_k}
\newcommand{\dbdchik}{\dot{\delta\bar{\chi}}_k}
\newcommand{\tzetak}{\tilde{\zeta}_k}
\newcommand{\tdzetak}{\dot{\tilde{\zeta}}_k}

\newcommand{\bphi}{\bar\phi}
\newcommand{\bchi}{\bar\chi}
\newcommand{\bX}{\bar X}
\newcommand{\bP}{\bar P}
\newcommand{\bY}{\bar Y}
\newcommand{\bH}{\bar H}
\newcommand{\bt}{\bar t}
\newcommand{\bF}{\bar F}
\newcommand{\bG}{\bar G}
\newcommand{\bA}{\bar A}
\newcommand{\bk}{\bar k}

\title{Fully viable DHOST bounce with extra scalar}
\author{Ok Song An,\,\,}
\emailAdd{os.an@ryongnamsan.edu.kp}
\author{Jin U Kang,\,\,}
\emailAdd{ju.kang0718@ryongnamsan.edu.kp}
\author{Yong Jin Kim,\,\,}
\emailAdd{cioc6@ryongnamsan.edu.kp}
\author{Ui Ri Mun\,\,}
\emailAdd{ur.mun0826@ryongnamsan.edu.kp}
\author{and\, Un Gyong Ri}
\emailAdd{uk.ri0107@ryongnamsan.edu.kp}

\affiliation{Department of Physics, \textbf{Kim Il Sung} University, Ryongnam Dong, Taesong District, Pyongyang, Democratic People's Republic of Korea}

\abstract{In this paper we construct a class of Degenerate Higher-Order Scalar-Tensor (DHOST) theories with an extra scalar field, which admits viable solutions of bouncing universe satisfying the following requirements: (i) absence of Belinski-Khalatnikov-Lifshitz (BKL) instability, ghost and gradient instability, (ii) absence of superluminality, (iii) generation of nearly scale-invariant curvature perturbations and very small tensor-to-scalar ratio, and (iv) conventional asymptotics in the distant past and future, where gravity sector is described by General Relativity and the DHOST scalar has a canonical form of Lagrangian. We also expect our models to have sufficiently small non-Gaussianities of primordial curvature perturbations to be compatible with observations. As such, this work exemplifies for the first time the fully viable two-field DHOST bouncing cosmology, which is free of instability and superluminality problems as well as compatible with observations.}
\preprint{}

\begin{document}
\maketitle

\section{Introduction}

The inflationary scenario \cite{Guth:1980zm,Starobinsky:1980te,Linde:1981mu} is the mainstream in studies of the early universe. Although it can solve many of the theoretical and phenomenological problems of the standard Big Bang cosmology, the initial singularity problem still remains unsolved \cite{Borde:2001nh}. Therefore, it deserves to study alternatives to the inflation.

The most promising alternative that can solve the initial singularity problem is a non-singular bouncing scenario (see \cite{Novello:2008ra,Battefeld:2014uga,Brandenberger:2016vhg} for a review), where the universe is supposed to undergo a smooth transition from the initial contracting phase to the currently expanding phase. In the non-singular bouncing scenario the Hubble parameter has to grow during the bounce period, which within the framework of General Relativity (GR) is possible only if there exists an exotic matter component that does not respect the Null Energy Condition (NEC). Here the problem is that it is challenging to realize the NEC violation without pathologies such as ghost and gradient instabilities and superluminality \cite{Rubakov:2014jja,Dubovsky:2005xd}.

The infamous no-go theorems \cite{Libanov:2016kfc,Kobayashi:2016xpl} tell us that in Horndeski theories \cite{Horndeski:1974wa} it is impossible for non-singular bouncing cosmologies not only to avoid ghost and gradient instabilities but also to have conventional asymptotics (so that strong coupling problem does not arise in the distant past and future). However, it turns out that beyond Horndeski \cite{Zumalacarregui:2013pma,Gleyzes:2014dya} and Degenerate Higher-Order Scalar-Tensor (DHOST) theories \cite{Langlois:2015cwa,BenAchour:2016cay,Langlois:2017mxy} (see \cite{Langlois:2018dxi,Kobayashi:2019hrl} for a review) admit non-singular bounce cosmology without ghost and gradient instabilities and with conventional asymptotics (see e.g. \cite{Creminelli:2016zwa,Kolevatov:2017voe,Ye:2019sth,Ye:2019frg,Ilyas:2020qja,Zhu:2021whu}).\footnote{Although they contain higher-derivative terms, these general scalar-tensor theories do not possess Ostrogradski instability by construction.}

An important yet unresolved issue is \textit{superluminality}, which is another undesirable feature of low energy effective field theory, since it indicates that such a theory can not be UV-completed in a Lorentz-covariant way \cite{Adams:2006sv}. Beyond-Horndeski theories admit non-singular bouncing scenario without ghost and gradient instabilities and superluminality and with conventional asymptotics \cite{Mironov:2018oec}. It was, however, shown in \cite{Mironov:2019mye,Mironov:2020mfo} that superluminality issue reappears once we add to the theory an extra scalar field minimally coupled to gravity which has the non-zero kinetic energy and the luminal sound speed. This superluminality problem can be resolved by either assuming \emph{strong gravity} in the asymptotic past and realizing the violation of NEC within Horndeski theories \cite{Ageeva:2018lko,Ageeva:2020buc,Ageeva:2020gti,Ageeva:2021yik,Ageeva:2022fyq,Ageeva:2022asq} or considering an exceptional subclass of DHOST Ia theories without the assumption of strong gravity as argued in \cite{Mironov:2020pqh,Mironov:2024xxx}. In general, strong gravity in the past causes a fast growth of quantum fluctuations, which implies that classical description of the model is no longer consistent. Although there exists a region in parameter space where this strong coupling problem does not occur \cite{Ageeva:2022fyq}, the bounce scenario with strong gravity in the past generates primordial scalar perturbations with large non-Gaussianities exceeding the observational constraints \cite{Ageeva:2024xxx}. On the other hand, to our best knowledge, no explicit model without superluminality and consistent with observations in DHOST Ia theories has been constructed so far.

The aim of this paper is to build a \emph{novel} explicit bouncing scenario within the exceptional subclass of DHOST Ia theories such that
\begin{itemize}
	\item   it is completely free from the aforementioned issues, i.e.,  ghost and gradient instabilities, superluminality without assuming strong gravity in the past and future,
	\item it is observationally viable in the sense that it predicts the power spectrum of perturbations compatible with observations.
\end{itemize} 
In order to seek such a healthy bouncing model within the exceptional subclass of DHOST Ia theories which does not suffer from any pathologies, we employ the inverse method used in \cite{Ijjas:2016tpn,Ijjas:2016vtq,Libanov:2016kfc,Kolevatov:2017voe,Cai:2017dyi,Mironov:2018oec}. In addition, to make the model observationally viable, we should introduce an extra scalar. The reasoning for this deserves a detailed discussion below.

One serious issue that has to be addressed in constructing a bounce scenario is the so-called Belinsky-Khalatnikov-Lifshitz (BKL) instability \cite{Belinsky:1970ew}, that is, the contracting phase is unstable due to the growth of anisotropies. In order to suppress the BKL instability within GR one needs to introduce a matter component with a steep equation of state that drives an ekpyrotic contraction phase (also known as a slow contraction phase) \cite{Khoury:2001wf,Buchbinder:2007ad,Levy:2015awa}, which turns out to be a robust and rapid smoother even for initial conditions far from being spatially-flat, homogeneous and isotropic \cite{Cook:2020oaj,Ijjas:2020dws,Ijjas:2021gkf,Ijjas:2024xxx}. In a word, the bounce scenario should begin with the ekpyrotic contraction phase in order for the BKL instability to be absent. 

The problem is that the ekpyrotic contraction makes all perturbations blue-tilted, which is incompatible with the observational data that the curvature perturbations have nearly scale-invariant power spectrum \cite{Aghanim:2018eyx}. A typical resolution to this issue is to utilize the entropic mechanism, which consists of two step: (i) adding an extra scalar field $ \chi $ that generates nearly scale-invariant perturbations during the ekpyrotic contraction phase and (ii) converting the entropic perturbations into curvature fluctuations around the bounce phase. It is now manifest that the healthy and phenomenologically viable bouncing scenario with conventional asymptotics should consist of two fields, where the adiabatic field $ \phi $ is described by the exceptional subclass of DHOST Ia theories.

As for the entropic field $ \chi $, we incorporate the entropic mechanism successfully with the exceptional subclass of DHOST Ia theories. Regarding the first step of the entropic mechanism, two classes of the scalar field $ \chi $ have been studied in the literature: the entropic field $ \chi $ has a canonical kinetic term and an unstable potential \cite{Notari:2002yc,Finelli:2002we,Lehners:2007ac}, or the entropic field $ \chi $ has a non-minimal kinetic coupling to the ekpyrotic scalar field \cite{Qiu:2013eoa,Li:2013hga,Fertig:2013yyy,Ijjas:2014fja}. In either case, the sound speed of $ \chi $ is equal to unity. The first class of models has a drawback that the potential has a tachyonic instability and thus the ekpyrotic contraction phase would not last long enough unless a special fine-tuned initial conditions are provided \cite{Buchbinder:2007tw}. On the other hand, the second class of models is stable and does not suffer from fine-tuning of initial conditions. For this reason, we choose the second class for generating a nearly scale-invariant power spectrum.

In the second step of the entropic mechanism the conversion can be done either by a repulsive potential \cite{Lehners:2007ac,Lehners:2009qu,Lehners:2010fy,Fertig:2015ola,Fertig:2016czu} or by a higher-order kinetic coupling of $ \chi $, where the sound speed of $ \chi $ becomes less than unity during the conversion process \cite{Ijjas:2020cyh,Ijjas:2021ewd}. In any case, the conversion is induced by a bending of the background trajectory in the field space. Here, the key requisite is that the conversion process should be efficient and smooth; otherwise the resulting curvature perturbations have a small amplitude and may have very large non-Gaussianities exceeding the current observational bounds, see \cite{Fertig:2013yyy,Fertig:2015ola,Fertig:2016czu} and references therein.

It is well-known that it is rather non-trivial to obtain an efficient conversion process \cite{Fertig:2015ola,Fertig:2016czu}, which becomes more challenging in our case due to the stability constraint. This is because the efficient conversion requires the total bending almost unity (see the argument of \cite{Fertig:2013yyy,Fertig:2015ola}), which can make the background trajectory get out of the stable region in the field space. In this paper we realize an efficient conversion process in terms of the entropic field with luminal sound speed. It must be noted, though, that all prior works related to conversion process have focused only on efficiency and non-Gaussianity but have not been concerned with the total background trajectory in the field space. In this paper we pay a careful attention to this point, namely we obtain an efficient conversion process maintaining the stability and non-superluminality throughout the whole background trajectory. By construction, our model is expected to have sufficiently small non-Gaussianities of primordial curvature perturbations. We again emphasize that this work exemplifies for the first time the fully viable two-field DHOST bouncing cosmology, which is free of instability and superluminality problems as well as compatible with observations.

The rest of this paper is organized as follows. In Section \ref{sec:model} we briefly review the DHOST Ia theories with an extra scalar field, in relation to the no-go theorem \cite{Mironov:2020pqh,Mironov:2024xxx}, and set up the theoretical framework and strategy employed in this paper. In Section \ref{sec:example} we consider the decoupled case (i.e., with unrolling extra scalar) and construct a stable and non-superluminal model in the DHOST sector. In Section \ref{sec:extra-field} we consider the coupled case (i.e., with rolling $ \chi $) and construct the extra scalar sector such that the stability and non-superluminality are not spoiled and the conversion from entropy to curvature perturbations is efficient enough to be compatible with observations. We summarize the result and conclude in Section \ref{sec:conclusion}. Throughout this paper we use the $ (+,-,-,-) $ signature of the metric and natural units with $ c=\hbar=1 $. We define the reduced Planck mass as $ M_{pl}\equiv 1/\sqrt{8\pi G_N}= 2.4\times 10^{18} $GeV.

\section{Exceptional Subclass of DHOST Ia with Extra Scalar}\label{sec:model}

In this section we briefly describe an exceptional subclass of  DHOST Ia theories with an extra scalar following the exposition in \cite{Mironov:2020pqh}, with a new ingredient related to the superluminality issue which motivates our work.

\subsection{DHOST Ia with extra scalar}

The action of DHOST Ia theories with an extra scalar field $\chi$ is given by
\begin{align}\label{eq:action}
\cS=\cS_\phi+\cS_\chi
\end{align}
where $ \cS_\phi $ is the quadratic action of the DHOST theory \cite{Langlois:2015cwa,Langlois:2017mxy}
\begin{subequations}
	\begin{align}\label{eq:action_phi}
	\mathcal S_\phi=\int d^4x\sqrt{-g}\left(F(\phi,X)+K(\phi,X)\Box\phi+ F_2(\phi,X)R+\sum_{I=1}^{5}A_I(\phi,X)L_I\right),
	\end{align}
	and $ \cS_\chi $ has the k-essence type \cite{ArmendarizPicon:2000ah}
	\begin{align}\label{eq:action_chi}
	\cS_\chi =\int d^4x\sqrt{-g}\;P(\chi,Y,\phi).
	\end{align}
\end{subequations}
Here $ X\equiv g^{\mu\nu}\phi_{,\mu}\phi_{,\nu} $, $ \phi_{,\mu}\equiv\partial_\mu\phi $, $ Y\equiv g^{\mu\nu}\chi_{,\mu}\chi_{,\nu}$, $\chi_{,\mu}\equiv\partial_\mu\chi $, and $ L_I $ ($ I=1,2,3,4,5 $) are
\begin{align*}
& L_1=\phi_{;\mu\nu}\phi^{;\mu\nu},\quad L_2=(\Box\phi)^2,\quad L_3=\phi^{,\mu}\phi_{;\mu\nu}\phi^{,\nu}\Box\phi,\\
& L_4=\phi^{,\mu}\phi_{;\mu\nu}\phi^{;\nu\rho}\phi_{,\rho},\quad L_5=(\phi^{,\mu}\phi_{;\mu\nu}\phi^{,\nu})^2,
\end{align*}
where $ \phi_{;\mu\nu}\equiv\nabla_\mu\nabla_\nu\phi $. The DHOST Ia theories are characterized by
\begin{subequations}
	\begin{align}
	A_2 =& -A_1,\\
	A_4 =& \frac{1}{8(F_2-XA_1)^2}\left[-16XA_1^3+4(3F_2+16XF_{2X})A_1^2-X^2F_2A_3^2 \right.\nonumber\\
	&\hskip2em -(16X^2F_{2X}-12XF_2)A_3A_1-16F_{2X}(3F_2+4XF_{2X})A_1\nonumber\\
	&\hskip2em\left.+8F_2(XF_{2X}-F_2)A_3+48F_2F_{2X}^2\right].\\
	A_5=&\frac{(4F_{2X}-2A_1+XA_3)(-2A_1^2-3XA_1A_3+4F_{2X}A_1+4F_2A_3)}{8(F_2-XA_1)^2},
	\end{align}
\end{subequations}
where the subscript $ X $ denotes the derivative with respect to $ X $. The independent functions of the model \eqref{eq:action} are then $ F(\phi,X) $, $ K(\phi,X) $, $ F_2(\phi,X) $, $ A_1(\phi,X) $, $ A_3(\phi,X) $ and $ P(\chi,Y,\phi) $. We emphasize that the coupling of $ \chi $ to the DHOST theory does not spoil the degeneracy, as the scalar field $ \chi $ does not feature higher derivatives and the coupling terms do not involve derivatives of $ \phi $ \cite{Deffayet:2020ypa,Mironov:2024xxx}. Hence, the whole system does not suffer from the Ostrogradsky ghost.

\subsection{Quadratic action for perturbations}

As mentioned in the Introduction, we aim to construct a bouncing scenario fully viable in the sense that (i) it is free of any pathologies such as ghost instability, gradient instability and superluminality and (ii) it predicts power spectra of primordial perturbations compatible with observations. On a spatially homogeneous background, the aforementioned pathologies, if present, manifest themselves in the linear perturbation theory, see e.g. \cite{Rubakov:2014jja}. Hence, we consider the linear theory of the model \eqref{eq:action}.

The perturbed spatially-flat Friedmann-Lema\^itre-Robertson-Walker (FLRW) metric is given by
\begin{align}\label{eq:perturbed-metric}
ds^2=(1+\alpha)^2dt^2-a^2(t)^{2\zeta}\left(\delta_{ij}+h_{ij}^{TT}+\frac12 h_{ik}^{TT} h_j^{k\;TT}\right)(dx^i+\partial^i\beta \;dt)(dx^j+\partial^j \beta \;dt),
\end{align}
where $ a(t) $ is the scale factor, $ \alpha, \beta $ and $ \zeta $ denote lapse, shift and curvature perturbations, respectively, and $ h_{ij}^{TT} $ is traceless and transverse tensor perturbation. We did not include the vector modes in the perturbed metric \eqref{eq:perturbed-metric}, as they are non-dynamical in scalar-tensor theories \cite{Mironov:2024xxx}.\footnote{Even though non-dynamical, the vector perturbations can be problematic in bouncing cosmologies, since they grow as $ 1/a^2 $ during the contracting phase \cite{Battefeld:2014uga}. The ekpyrotic contracting phase, which is employed in this work, smooths and flattens the universe, since the energy density of the homogeneous ekpyrotic scalar field grows faster than $ 1/a^6 $, so that the vector perturbations remain negligible compared to the background. The regrowth of vector modes during the bounce phase discussed in \cite{Xue:2011nw} does not occur in this work, because the overall change in the scale factor during the bounce phase is negligible, see \eqref{eq:scale-factor} and Fig. \ref{fig:ah}.} Throughout this paper we work in the unitary gauge $ \delta\phi=0 $. The lapse and shift perturbations $ \alpha$ and $\beta $ appear in the quadratic action as Lagrange multipliers, so they can be integrated out. After all, we obtain a quadratic action $ S^{(2)}=S^{(2)}_{\text{tensor}}+S^{(2)}_{\text{scalar}} $ for the perturbations  $h_{ij}^{TT} $, $ \zeta$ and $ \delta\chi $, where
\begin{align}\label{eq:quadratic-action-tensor}
S^{(2)}_{\text{tensor}}=\int dt\;d^3x\;a^3\left(\frac{\mathcal G_{\mathcal T}}{8}(\dot h_{ij}^{TT})^2-\frac{\mathcal F_{\mathcal T}}{8a^2}(\partial_k h_{ij}^{TT})^2\right),
\end{align}
with
\begin{subequations}
	\begin{align}
	\mathcal G_{\mathcal T} = &\; -2F_2+2A_1 X,\\
	\mathcal F_{\mathcal T} = &\; -2F_2.
	\end{align}
\end{subequations}
Here the dot notation denotes the derivative with respect to time $ t $. Note that perturbation $ \delta\chi $ of the extra scalar $ \chi $ does not appear in the quadratic action for  tensor perturbations, but appears only in the quadratic action $ S^{(2)}_{\text{scalar}} $ for the scalar perturbations $ \zeta $ and $ \delta\chi $ given by \cite{Mironov:2020pqh,Mironov:2024xxx}
\begin{align}\label{eq:quadratic-action-perturbations}
S^{(2)}_{\text{scalar}}=\int dt\;d^3x\;a^3 \left(G_{AB}\dot v^A \dot v^B-\frac{1}{a^2}F_{AB}\delta^{ij}\partial_i v^A\partial_j v^B+\Psi_1\dot{\tilde\zeta}\delta\chi+\Psi_2\delta\chi^2 \right).
\end{align}
Here $ A,B=1,2 $, and $ v^1$ and $ v^2 $ denote $\tilde\zeta\equiv \zeta+\frac{\Gamma}{2\mathcal G_{\mathcal T}}\alpha $ and $ \delta\chi $, respectively, where
\begin{align}
\Gamma = X(-2A_1+4F_{2X}+A_3X).
\end{align}
The functions $ \Psi_{1,2} $ and the matrices $ G_{AB} $ and $ F_{AB} $ are given by respectively
\begin{subequations}
	\begin{align}
	\Psi_1=&\;\frac{\mathcal G_{\mathcal T}}{\Theta^2}\left[2(YQ+\Sigma)\dot\chi P_Y+\Theta\left(P_\chi-2YP_{\chi Y}+\frac{6\Delta}{a^3}\frac{d}{dt}\left(a^3\dot\chi P_Y\right) \right) \right] ,\\
	\Psi_2=&\;\Omega+\frac{1}{a^3}\frac{d}{dt}\left(\frac{a^3 YP_Y Q\left(1-3\frac{P_Y}{Q}\Delta\right)}{\Theta}\right)\nonumber\\
	&\hskip1em+\frac{\dot\chi P_Y}{\Theta^2}\left[\dot\chi P_Y(YQ+\Sigma)+\Theta\left(P_\chi-2YP_{\chi Y}+\frac{6\Delta}{a^3}\frac{d}{dt}\left(a^3\dot\chi P_Y\right) \right) \right] ,\\
	G_{AB}= &\;\begin{pmatrix}
	\GS+\frac{\GT^2}{\Theta^2}YQ & \dot\chi Qg\\
	\dot\chi Qg & Q
	\end{pmatrix},\quad F_{AB}=\begin{pmatrix}
	\FS & \dot\chi P_Y f\\
	\dot\chi P_Y f & P_Y
	\end{pmatrix},\label{kinetic matrix}
	\end{align}
\end{subequations}
where the subscripts $ \chi $ and $ Y $ denote derivative with respect to $ \chi $ and $ Y $, respectively. Notice that in the case of $ \dot\chi=0 $ the matrix $ G_{AB} $ and $ F_{AB} $ become diagonal so that the DHOST sector decouples from $ \chi $ sector. We also introduced the new functions \footnote{We note that the expression for $ \Sigma $ given in \cite{Mironov:2020pqh} has some typos; e.g., the prefactor of $ \dot\phi^3 $ should not contain $ \dddot\phi $. In this paper $ \Sigma $ and $ \Theta $ correspond to $ \tilde\Sigma $ and $ \tilde\Theta $ in \cite{Mironov:2020pqh,Mironov:2024xxx}, respectively.}
\begin{subequations}
	\begin{align}
	\Omega=&\;\frac12 P_{\chi\chi}-\frac{1}{a^3}\frac{d}{dt}\left(a^3\dot\chi P_{\chi Y}\right),\\
	Q = &\;P_Y+2Y P_{YY},\\
	f=&\;-\frac{\mathcal G_{\mathcal T}+\mathcal D\dot\phi+\mathcal F_{\mathcal T}\Delta}{\Theta},\\
	g =&\; -\frac{\mathcal G_{\mathcal T}}{\Theta}\left(1-3\frac{P_Y}{Q}\Delta\right),\\
	\mathcal G_{\mathcal S}= &\;\frac{\Sigma \mathcal G_{\mathcal T}^2}{\Theta^2}+3\mathcal G_{\mathcal T} ,\label{eq:GS}\\
	\mathcal F_{\mathcal S}= &\;\frac1a \frac{d}{dt}\left(\frac{a\mathcal G_{\mathcal T}(\mathcal G_{\mathcal T}+\mathcal D\dot\phi+\mathcal F_{\mathcal T}\Delta)}{\Theta}\right)-\mathcal F_{\mathcal T} ,\label{eq:FS}\\
	\mathcal D = &\; 2\dot\phi(F_{2X}-A_1),\\
	\Delta =&\;\frac{X}{2\mathcal G_{\mathcal T}}(2A_1-4F_{2X}-A_3X),\\
	\Theta = &\; -\mathcal G_{\mathcal T}\dot\Delta-2 F_2 H+\dot\phi \left(3 \ddot\phi \left(A_1-2 F{2X}\right)-F_{2\phi }\right)+H \dot\phi^2 \left(3 A_1+2 F_{2X}\right)\nonumber\\
	& +\dot\phi^3 \left(\frac{1}{2} \ddot\phi \left(4 A_{1X}-3 A_3+2 A_4-8 F_{2XX}\right)-2 F_{2\phi X}-K_X\right)\nonumber\\
	&+\frac{1}{2} H \dot\phi^4 \left(8 A_{1X}+3 A_3\right)+\dot\phi^5\ddot\phi \left(A_5-A_{3X}\right)\label{eq:tilde-Theta},\\
	\Sigma = &\;3\GT\dot\Delta^2+6\Theta\dot\Delta-\frac{3}{a^3}\frac{d}{dt}\left[a^3(\Theta+\GT\dot\Delta)\Delta\right]\nonumber\\
	&+ 6 F_2 H^2+6 H \dot\phi \left(F_{2\phi}-2 \left(A_1-2 F_{2X}\right) \ddot\phi\right)\nonumber\\
	&+\dot\phi^2 \left(12 A_1 \dot H-3 A_3 \ddot\phi^2-3 A_4 \ddot\phi^2+F_X-24 F_{2X} \dot H-42 F_{2X} H^2-K_{\phi }\right)\nonumber\\
	&-3 \dot\phi^3 \left(H \left(2 A_{1X} \ddot\phi-4 A_{1\phi}+3 A_3 \ddot\phi+6 A_4 \ddot\phi-4 F_{2XX} \ddot\phi-2 F_{2\phi X}-4 K_X\right)+2 (A_3+A_4) \dddot\phi\right)\nonumber\\
	&+ \dot\phi^4 \left(-3 H^2 \left(12 A_{1X}+9 A_3+8 F_{2XX}\right)+6 A_{1X} \dot H-9 A_3 \dot H-9 A_{3X}\ddot\phi^2-6 A_{3\phi} \ddot\phi-9 A_{4X} \ddot\phi^2\right.\nonumber\\
	&\hskip4em\left.-6 A_{4\phi} \ddot\phi-12 A_5 \ddot\phi^2+2 F_{XX}-12 F_{2XX} \dot H-K_{\phi X}\right)\nonumber\\
	&+\dot\phi^5 \left(3 H \left(2 A_{1\phi X}-A_{3X}\ddot\phi-3 A_{3\phi}-2 A_{4X} \ddot\phi-8 A_5 \ddot\phi+2 K_{\text{XX}}\right)-2 \dddot\phi \left(A_{3X}+A_{4X}+4 A_5\right)\right)\nonumber\\
	&+\dot\phi^6 \left[-3 H^2 \left(4 A_{1XX}+3 A_{3X}\right)-3 A_{3X}\dot H\right.\nonumber\\
	&\left.\hskip4em-\ddot\phi \left(2 A_{3XX} \ddot\phi+2 A_{3\phi X}+2 A_{4XX} \ddot\phi+2 A_{4\phi X}+13 A_{5X}\ddot\phi+8 A_{5\phi}\right)\right]\nonumber\\
	&+\dot\phi^7 \left(-3 H \left(A_{3\phi X}+2 A_{5X}\ddot\phi\right)-2 A_{5X}\dddot\phi\right)-2 \dot\phi^8 \ddot\phi \left(A_{5XX} \ddot\phi+A_{5\phi X}\right).\label{eq:Sigma}
	\end{align}
\end{subequations}
Notice that terms with $ \zeta^2 $ do not appear in the quadratic action $ S^{(2)}_{\text{scalar}} $ since they cancel out upon using background equations of motion.

\subsection{Stabilities and non-superluminality conditions: an exceptional subclass}

The main types of instabilities that can potentially emerge at the linearized level of perturbations in the high-momentum regime are ghost and gradient instabilities (see \cite{Rubakov:2014jja} for details). The presence of a ghost instability implies that the vacuum is quantum mechanically unstable, while the gradient instability leads to the exponential growth of perturbations.

For the tensor modes, the ghost instability appears when $ \GT<0 $ and $ \FT<0 $ in \eqref{eq:quadratic-action-tensor}, while the gradient instability shows up when $ \GT>0 $, $ \FT<0 $ or $ \GT<0 $, $ \FT>0 $. The stable situation thus corresponds to $ \GT>0 $ and $ \FT>0 $. On the other hand, the non-superluminality condition, which requires the propagation speed of tensor perturbations not greater than unity, is $ c_T^2\equiv \FT/\GT\leq1 $. Thus, the stability and non-superluminality conditions for the tensor sector are given by
	\begin{align}\label{eq:stability-condition-tensor}
	\GT\geq\FT>0.
	\end{align}

Similar arguments apply to the scalar sector, but two-field cases require some involved algebra. First, the stability conditions amount to the positive definiteness of kinetic matrices \eqref{kinetic matrix} (see \cite{Mironov:2020pqh,Mironov:2024xxx}), i.e.
	\begin{subequations}\label{eq:stability-condition-scalar}
		\begin{align}
		& Q>0,\quad \Pi\equiv \Theta^2\GS+\left(\GT^2-\Theta^2g^2\right)YQ>0,\\
		& P_Y>0,\quad \Xi\equiv \FS-YP_Y f^2>0.
		\end{align}
	\end{subequations}
Second, to find the non-superluminality condition we compute the squared sound speeds, which are the eigenvalues of the matrix $G^{-1}F$
\begin{align}\label{eq:squared-sound-speed}
c_{s,\pm}^2=\frac{1}{2 \Pi }\left(\cA\pm \sqrt{\cA^2-4 \Xi  \Pi  \Theta^2 c_m^2}\right),
\end{align}
where
\begin{align}
\cA=&\;\Pi c_m^2+\Theta^2 Q Y c_m^2 (f-g)^2+\Theta^2\Xi,
\end{align}
and $ c_m^2=P_Y/Q $ is the squared sound speed of the extra scalar field $ \chi $. Since $ c_{s,+}^2\geq c_{s,-}^2 $ from \eqref{eq:squared-sound-speed}, it is enough to consider only $ c_{s,+}^2 $ as long as we are concerned with the superluminality issue.  Rewriting \eqref{eq:squared-sound-speed} as
\begin{align}\label{eq:squared-sound-speed-re}
c_{s,+}^2=&\;\frac{1}{2 \Pi }\left(\Pi c_m^2+\Theta^2 Q Y c_m^2 (f-g)^2+\Theta^2\Xi \right.\nn\\
&\left.+ \sqrt{\left(\Pi c_m^2+\Theta^2 Q Y c_m^2 (f-g)^2-\Theta^2\Xi\right)^2+4\Theta^4 QYc_m^2\Xi (f-g)^2}\right),
\end{align}
it follows from the stability condition \eqref{eq:stability-condition-scalar} that
\begin{align}\label{eq:cs+}
c_{s,+}^2\geq c_m^2+\frac{\Theta^2 Q Y c_m^2 (f-g)^2}{\Pi}\geq c_m^2.
\end{align}

In the case that the extra scalar field has luminal sound speed, i.e. $ c_m^2=1 $, the non-superluminality condition $ c_{s,+}^2\leq 1 $ combined with the inequality \eqref{eq:cs+} implies
\begin{align}\label{eq:Exceptional-A3}
f=g\quad \text{or }\quad A_3=\frac{2 (X A_1-2 F_2) \left(A_1-2 F_{2X}\right)}{X (3 X A_1-4 F_2)}.
\end{align}
The model that satisfies the above condition is dubbed \emph{exceptional subclass} of DHOST Ia theories in \cite{Mironov:2020pqh,Mironov:2024xxx}.\footnote{As shown here, the condition \eqref{eq:Exceptional-A3} holds for any value of $ Y $. In \cite{Mironov:2020pqh,Mironov:2024xxx} the emergent superluminality was analyzed for small $ Y $.} We emphasize that DHOST Ia theories other than the exceptional subclass necessarily lead to superluminality, when combined with a luminal extra scalar.

\subsection{Sufficient conditions for non-superluminality}

The exceptional subclass may avoid the superluminaltity problem, but it is still not completely free of superluminality, since the condition \eqref{eq:Exceptional-A3} is necessary, but not sufficient for non-superluminality. In other words, we may have $ c_{s,+}^2>1 $ even though the condition \eqref{eq:Exceptional-A3} is satisfied.

Indeed, when $ f=g $, the squared sound speed $ c_{s,+}^2 $ is reduced to
\begin{align}
c_{s,+}^2=&\;\frac{1}{2 \Pi }\left(\Pi c_m^2+\Theta^2\Xi + \left|\Pi c_m^2-\Theta^2\Xi\right|\right),
\end{align}
which becomes for $ \Pi c_m^2<\Theta^2\Xi $ that
\begin{align}
c_{s,+}^2=\frac{\Theta^2\Xi}{\Pi}=\frac{\Theta^2(\FS-YP_Y g^2)}{\Theta^2\GS+\left(\GT^2-\Theta^2 g^2\right)Y Q}.
\end{align}
In the case of $ \Theta^2\Xi>\Pi$ one can readily see that $ c_{s,+}^2>1$ and thus the superluminality appears in the scalar sector. Therefore, we find that the sufficient condition for the non-superluminality should be augmented by $ \Theta^2\Xi<\Pi $, which is achieved once $ c_m^2=1 $ and 
\begin{align}\label{eq:FS-YPYg2}
\FS>YP_Y g^2.
\end{align}

To sum up, the  sufficient conditions to have the non-superluminal sound speed in the exceptional subclass of DHOST Ia theories are: (i) the extra scalar field has the luminal sound speed, (ii) the stability conditions \eqref{eq:stability-condition-tensor} and \eqref{eq:stability-condition-scalar} holds, (iii) the scalar sector of the original DHOST action is non-superluminal (i.e., $ \GS\geq \FS $), (iv) the newly augmented sufficient condition \eqref{eq:FS-YPYg2} is satisfied. All of these can be combined into the simple conditions
\begin{align}\label{eq:stability-condition-scalar-final}
\GT\geq\FT>0,\quad \GS\geq\FS>YP_Y g^2.
\end{align}

In the subsequent sections we will construct an explicit bouncing scenario free of pathologies (that is, satisfying the constraint \eqref{eq:stability-condition-scalar-final} together with \eqref{eq:stability-condition-tensor}) as well as phenomenologically viable. The roadmap for this is as follows:
\begin{enumerate}
	\item (Section \ref{sec:example}) assuming that $ \chi $ does not roll (i.e., $ Y=0 $) so that the scalar perturbations $ \tilde\zeta $ and $ \delta\chi $ decouple from each other, we construct a completely healthy action $ \cS_\phi $ in the exceptional subclass of DHOST Ia theories, in the sense that it is free from
	\begin{itemize}
		\item BKL instability,
		\item strong gravity assumption, in other words described both in the distant past and future by Einstein gravity and a conventional scalar field with luminal sound speed,
		\item superluminality as well as ghost and gradient instability, i.e. obeys the conditions \eqref{eq:stability-condition-tensor} and \eqref{eq:stability-condition-scalar-final} with $ Y=0 $
		\begin{align}\label{eq:phi-constraint}
		\mathcal G_{\mathcal T}\geq \mathcal F_{\mathcal T}>0,\quad \mathcal G_{\mathcal S}\geq \mathcal F_{\mathcal S}>0,
		\end{align}
	\end{itemize}
	\item (Section \ref{sec:extra-field}) allowing that the extra scalar $ \chi $ rolls and interacts with $ \phi $, we construct the action $ \cS_\chi $ such that
	\begin{itemize}
		\item $ \chi $ has luminal sound speed,
		\item during the ekpyrotic contraction phase a nearly scale-invariant power spectrum for $ \delta\chi $ is generated,
		\item the perturbation $ \delta\chi $ is converted into the curvature perturbation $ \tilde\zeta $ after the bounce,
		\item while the original stability and non-superluminality of $ \cS_\phi $ are not spoiled, i.e.
		\begin{align}
		&\FS>YP_Y g^2.\label{eq:positivity-FAB}
		\end{align}
	\end{itemize}
\end{enumerate}

\section{DHOST Sector}\label{sec:example}

In this section we consider the decoupled case (i.e., $ \dot\chi=0 $) to construct a fully healthy model in the exceptional subclass of DHOST Ia, in which the independent Lagrangian functions are $ F(\phi,X) $, $ F_2(\phi,X) $, $ K(\phi,X) $ and $ A_1(\phi,X) $. We remind that the Lagrangian function $ A_3(\phi,X) $ is no longer independent in the exceptional subclass due to the condition \eqref{eq:Exceptional-A3}. To this end, we adopt the \textit{inverse method} used in \cite{Ijjas:2016tpn,Ijjas:2016vtq,Libanov:2016kfc,Kolevatov:2017voe,Cai:2017dyi,Mironov:2018oec} to choose a specific form of the Hubble parameter $ H(t) $ and reconstruct the Lagrangian functions $ F(\phi(t),X(t))$, $ F_2(\phi(t),X(t))$, $ K(\phi(t),X(t))$ and $ A_1(\phi(t),X(t)) $ that obey the background equations of motion and satisfy the constraint \eqref{eq:phi-constraint}.

Let us suppose that the DHOST scalar field $ \phi $ is a monotonic function of cosmic time $ t $. Then, without loss of generality we can choose the form of $ \phi $ as follows
\begin{align}\label{eq:phi}
\phi(t)=\mu^2 t,
\end{align}
where $ \mu $ is a characteristic energy scale. This form of $ \phi $ will greatly simplify the DHOST-related equations. Note that above form of $ \phi $ is related to the scalar field $ \varphi $ with asymptotically canonical kinetic term (i.e. the coefficient of the kinetic term $ \varphi_\mu\varphi^\mu $ approaching to $ 1/2 $ asymptotically) by $ \phi\sim\exp(\varphi) $ \cite{Mironov:2018oec}.

It is convenient to work on the rescaled quantities:
\begin{subequations}\label{eq:rescaled-phi}
	\begin{align}
	& t=\frac{M_{pl}}{\mu^2}\bar t,\quad H=\frac{\mu^2}{M_{pl}}\bar H,\quad \phi=M_{pl}\bar\phi,\quad X=\mu^4\bar X,\label{eq:time-dimensionless}\\
	& F=\mu^4 \bar F,\quad F_2=M_{pl}^2\bar F_2,\quad K=\mu \bar K, \quad A_1=\frac{M_{pl}^2}{\mu^4}\bar A_1,\\
	& \Theta=\mu^2 M_{pl}\btTh,\quad \Sigma=\mu^4\btSig,\quad \mathcal D=\frac{M_{pl}^2}{\mu^2}\bcD, \\
	& \GT=M_{pl}^2\bGT,\quad \FT=M_{pl}^2\bFT,\quad \GS=M_{pl}^2\bGS,\quad \FS=M_{pl}^2\bFS,
	\end{align}
\end{subequations}
where the quantities with an overbar are dimensionless. Note that the background solution \eqref{eq:phi} corresponds to $ \bphi(\bt)=\bt $ from \eqref{eq:time-dimensionless}, so that $ \bt $ will be replaced by $ \bphi $ at the end of the reconstruction procedure to obtain the Lagrangian functions as functions of the field $ \phi $. From now on, the dot notation denotes the derivative with respect to the dimensionless time $ \bt $.

\subsection{Asymptotics and ansatz}

Since in our model gravity and the scalar field $ \phi $ become conventional in the asymptotic limit, the model functions $\bar F_2(\bphi,\bX) $, $\bar K(\bphi,\bX)  $ and $\bar A_I(\bphi,\bX) $ ($ I=1,2,3,4,5 $) should have the following asymptotic behavior:
\begin{align}\label{eq:asymptotics-F2-K-AI}
\bF_2(\bphi,\bX)\sim -\frac12,\quad \bar K(\bphi,\bX),\; \bA_I(\bphi,\bX)\sim 0,\; \text{for }|\bt|\to+\infty.
\end{align}

The asymptotic behavior of the rescaled Hubble parameter $ \bH $ and the rescaled model function $ \bF(\bphi,\bX) $ can be different in the far past and in the far future. We choose the contracting phase to be ekpyrotic in order to suppress the BKL instability \cite{Belinsky:1970ew}; otherwise anisotropies can grow and dominate, eventually spoiling the isotropy of FLRW cosmology and inducing a chaotic mixmaster phenomenon. Moreover, the ekpyrotic contraction is a rapid and robust smoother, meaning that homogeneity, isotropy and flatness are achieved within a few e-folds even in the case that the initial state is far from homogeneous, isotropic and flat universe \cite{Cook:2020oaj,Ijjas:2020dws}.

During the ekpyrotic contraction phase, the dimensionless Hubble parameter $ \bH $ and the model function $ \bF(\bphi,\bX) $ behave as $ \bH\sim \frac{1}{\epsilon \bt} $, $ \bF(\bphi,\bX)\sim -2\dot \bH-3\bH^2\sim\frac{2\epsilon-3}{\epsilon^2 \bt^2} $ with $ \epsilon>3 $ as $ \bt\to-\infty $. On the other hand, we let the rescaled Hubble parameter $ \bH\sim \frac{1}{3\bt} $ and the model function $ \bF(\bphi,\bX)\sim \frac{1}{3\bt^2} $ as $ \bt\to+\infty $ in the expanding phase as in \cite{Mironov:2018oec}. This solution corresponds to the kinetic-dominated expanding universe, which is in fact a scaling attractor when the universe is filled only with a scalar field with sufficiently steep exponential potential \cite{Heard:2002dr}.  In total, we let the scale factor $ a(\bt) $, the Hubble parameter $ \bH(\bt)\equiv \dot a/a $ and $ \bF(\bphi,\bX) $ behave asymptotically as follows:
\begin{align}\label{eq:asymptotics-H-F}
a\propto \begin{cases}
\bt^{\frac 13}, & \bt\to+\infty\\
(-\bt)^{\frac 1\epsilon}, & \bt\to-\infty
\end{cases},\quad \bH\sim \begin{cases}
\frac{1}{3\bt}, & \bt\to+\infty\\
\frac{1}{\epsilon \bt}, & \bt\to-\infty
\end{cases},\quad \bF(\bphi,\bX)\sim\begin{cases}
\frac{1}{3\bt^2}, & \bt\to+\infty\\
\frac{2\epsilon-3}{\epsilon^2 \bt^2} , & \bt\to-\infty
\end{cases}.
\end{align}

\begin{figure}[t]
	\centering
	\includegraphics[width=\linewidth]{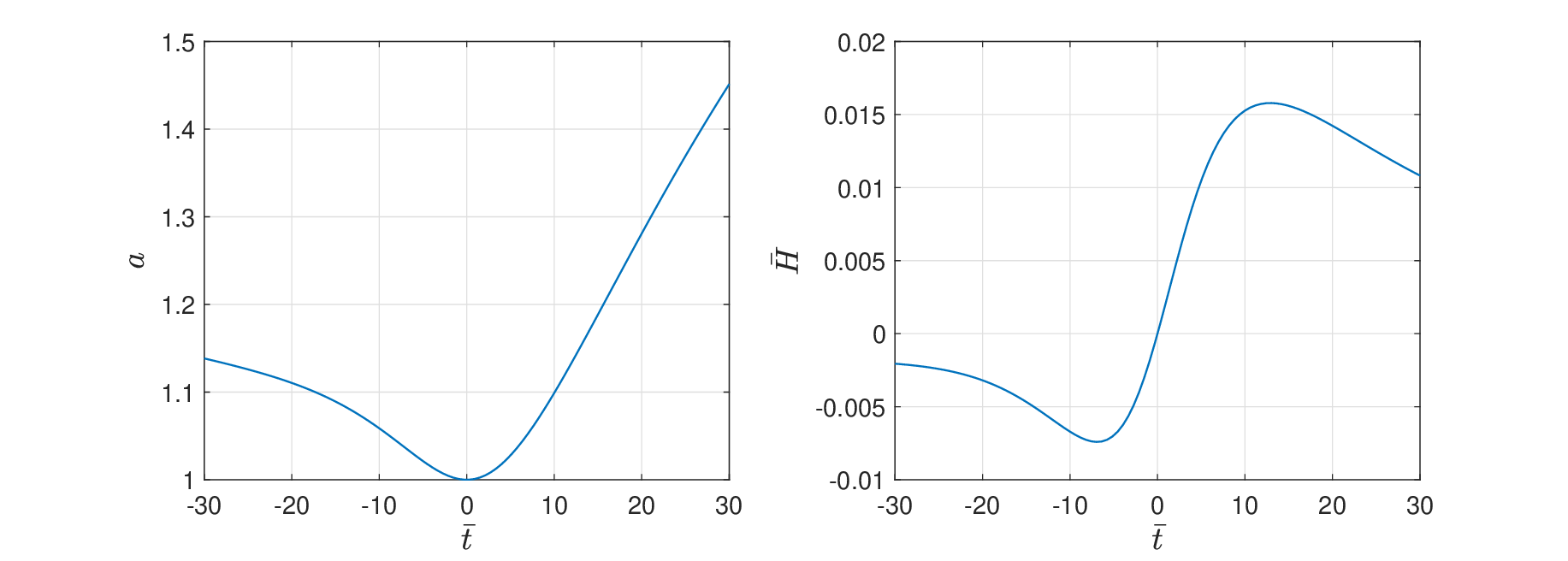}
	\caption{Plot of the scale factor $ a $ and the rescaled Hubble parameter $ \bH $. The parameters $ \tau $ and $ \epsilon $ are set to be $ \tau=10 $ and $ \epsilon=10 $.}
	\label{fig:ah}
\end{figure}

Our choice for $ a(\bt) $ with the above asymptotic behavior is
\begin{align}\label{eq:scale-factor}
a(\bt)=\sigma \left(\frac{\bt}{\tau }\right)\left[\left(\frac{\bt}{\tau }\right)^2+1\right]^{\frac 16}+ \sigma \left(-\frac{\bt}{\tau }\right) \left[\left(\frac{\bt}{\tau }\right)^2+1\right]^{\frac{1}{2 \epsilon }},
\end{align}
where $ \sigma(\bt)\equiv 1/(1+\exp(-\bt)) $ is the logistic sigmoid function, and $ \tau>0 $ is a parameter that controls the duration of the bounce phase. For this choice, the bounce occurs exactly at $ \bt=0 $, see Fig. \ref{fig:ah}. Note that ignoring the exponentially suppressed terms, the rescaled Hubble parameter $ \bH $ behaves as
\begin{align}
\bH(\bt)\sim \frac{\bt}{\gamma(\bt^2+\tau^2)}\quad \text{ for } |\bt|\to+\infty,
\end{align}
where $ \gamma=3 $ for $ \bt>0 $ and $ \gamma=\epsilon $ for $ \bt<0 $.

We make an ansatz for the independent Lagrangian functions $ \bF(\bphi,\bX) $, $ \bar K(\bphi,\bX) $, $ \bF_2(\bphi,\bX) $ and $ \bA_1(\bphi,\bX) $ as follows:
\begin{subequations}\label{eq:ansatz}
	\begin{align}
	\bF(\bphi,\bX) & \; = f_0(\bphi)+f_1(\bphi)\bX+f_2(\bphi)\bX^2,\\
	\bF_2(\bphi,\bX) & \; = -\frac12 -g_0(\bphi)-g_1(\bphi)\bX,\\
	\bA_1(\bphi,\bX) & \; = a_0(\bphi)+a_1(\bphi)\bX,\\
	\bar K(\bphi, \bX) & \; = 0.
	\end{align}
\end{subequations}
We let the functions $ g_{0,1}(\bt) $ and $ a_{0,1}(\bt) $ be exponentially suppressed as $ |t|\to+\infty $, i.e.{\footnote{The functions $a_{0,1}(\bar{t})$ should not be confused with the scale factor $a(\bt)$.}
\begin{align}\label{eq:asymptotics-a-g}
g_{0,1}(\bt), a_{0,1}(\bt)\propto e^{-\frac{2|\bt|}{\tau}}\quad \text{for } |\bt|\to+\infty.
\end{align}
During the ekpyrotic contraction phase ($ \bt<0 $), $ f_0(\bt) $ and $ f_1(\bt) $ behave as
\begin{align}\label{eq:asymptotics-f0-f1-past}
f_0(\bt)\sim \frac{\epsilon-3}{\epsilon^2\bt^2},\quad f_1(\bt)\sim \frac{1}{\epsilon \bt^2} \quad \text{ for } \bt\to-\infty ,
\end{align}
while during the (kinetic-dominated) expanding phase ($ \bt>0 $) the potential term $ f_0 $ becomes subdominant, namely
\begin{align}\label{eq:asymptotisc-f0-f1-future}
f_0(\bt)\propto \frac{1}{\bt^4},\quad f_1(\bt)\sim \frac{1}{3 \bt^2}\quad \text{ for } \bt\to+\infty.
\end{align}
The function $ f_2(\bt) $ should be subdominant compared to $ f_1(\bt) $, and therefore $ f_2(\bt) $ should behave asymptotically as follows:
\begin{align}\label{eq:asymptotics-f2}
f_2(\bt)\propto \frac{1}{\bt^4}\quad \text{for } |\bt|\to+\infty.
\end{align}

\subsection{Stability and non-superluminality}

Now we choose functions $ f_{0,1,2}(\bt) $, $ g_{0,1}(\bt) $ and $ a_{0,1}(\bt) $ such that not only the background equations of motion but also the constraint \eqref{eq:phi-constraint} is satisfied.

First, the constraint \eqref{eq:stability-condition-tensor} for the tensor sector is easy to be satisfied. Indeed, for the ansatz \eqref{eq:ansatz}, the coefficients of the quadratic action in the tensor sector become
\begin{align}
\bGT=1+ 2\left[a_0(\bt)+a_1(\bt)+g_0(\bt)+g_1(\bt)\right],\quad \bFT=1+2\left[g_0(\bt)+g_1(\bt)\right].
\end{align}
We select $ g_0 $ and $ a_0 $ such that $ \bGT=\bFT=1 $, namely
\begin{align}\label{eq:g0-a0}
g_0(\bt)=-g_1(\bt),\quad a_0(\bt)=-a_1(\bt).
\end{align}
That is, the speed of gravitational wave is always equal to unity and all the pathologies are absent in the tensor sector.

Now we determine the functions $ f_{0,1,2}(\bt) $, $ a_1(\bt) $ and $ g_1(\bt) $ such that none of the pathologies appear in the scalar sector, i.e. $ \bGS\geq\bFS>0 $. Notice that only $ a_1(\bt) $ and $ g_1(\bt) $ among these five Lagrangian functions appear in $ \bFS $, namely
\begin{align}
\bFS=\frac{1}{a}\frac{d\xi}{d\bt}-1,
\end{align}
where
\begin{align}
\xi\equiv \frac{a\Lambda}{\btTh},
\end{align}
with
\begin{align}
\Lambda\equiv&\; \bGT+\bcD\dot{\bar{\phi }}+\bFT\Delta=1-3 g_1(\bt),\\
\btTh=&\;\bar H \left(4 a_1(\bt)+g_1(\bt)+1\right)+\dot g_1(\bt).
\end{align}
We therefore take the first step to choose $ a_1(\bt) $ and $ g_1(\bt) $ such that $ \bFS>0 $.

Since $ \dot\xi=a\left(1+\bFS\right)>0$, one can naively conclude that $ \xi $ should be a monotonically increasing function. However, one must be careful since $ \btTh\sim \bar H $ as $ |\bt|\to+\infty $ and thus $ \btTh $ should cross zero at least once. Indeed, both $ \bGS $ and $ \bFS $ become divergent when $ \btTh=0 $. As shown in \cite{Mironov:2018oec}, this is entirely harmless, because the equation of motion for $ \tilde \zeta $ has regular solutions. A similar situation was discussed in \cite{Koehn:2015vvy}.

We assume that $ \btTh $ crosses zero only once for simplicity and denote the time when $ \btTh $ vanishes by $ \bt_0 $. Then, the function $ \xi(\bt) $ should increase monotonically except for the moment $ \bt=\bt_0 $. In particular, $ \xi(\bt) $ should behave in the vicinity of $ \bt=\bt_0 $ as
\begin{align}
\lim_{\bt\to \bt_{0-}}\xi(\bt)=+\infty,\quad \lim_{\bt\to \bt_{0+}}\xi(\bt)=-\infty.
\end{align}
On the other hand, it follows from the form of $ \bH $ and the asymptotics \eqref{eq:asymptotics-a-g} that $ \xi(-\infty)<0 $ and $ \xi(+\infty)>0 $. Therefore, $ \Lambda $ crosses zero once in the interval $ (-\infty,\bt_0) $ and $ (\bt_0,+\infty) $, respectively.

To achieve this, we choose the form of $ g_1(\bt) $ as follows:
\begin{align}\label{eq:g1}
g_1(\bt)=\frac{1}{3} w \cosh^{-2}\left(\frac{\bt}{\tau }+u\right),
\end{align}
where the parameter $ \tau $ controls the duration of the bounce phase. The parameters $ w>1 $ and $ u $ are introduced to avoid the fine-tuning problem that $ \bH $, $ \btTh $ and $ \xi $ become zero at the same time \cite{Mironov:2018oec}. We also choose $ a_1(\bt) $ such that $\btTh $ is sufficiently close to $ \bH $ for all times (see Fig. \ref{fig:thetatilde}).
\begin{align}\label{eq:a1}
a_1(\bt)=\frac{w}{12}\text{sech}^2\left(\frac{\bt}{\tau }+u\right) \left(\frac{2 \left(\bt^2 \tanh \left(\frac{\bt}{\tau }+u\right)+\tau ^2 \tanh \left(\frac{\bt}{\tau }\right)\right)}{\bH \tau  \left(\bt^2+\tau ^2\right)}-1\right) .
\end{align}
Then, $ \btTh $ becomes
\begin{align}
\btTh=&\;\frac{1}{3 \tau  \epsilon }\left[\frac{3 \bt \tau }{\bt^2+\tau ^2}+\frac{e^{\bt/\tau } \left(3 \bt^2 \epsilon +\tau  \bt (\epsilon -3)+3 \tau ^2 \epsilon \right)}{e^{\bt/\tau } \left(\bt^2+\tau ^2\right)+\tau ^2 \left(\frac{\bt^2}{\tau ^2}+1\right)^{\frac{1}{2 \epsilon }+\frac{5}{6}}}+3 \epsilon  \left(\frac{1}{e^{\bt/\tau }+1}-1\right)\right]\nonumber\\
&\hskip5em-\frac{2 \tau  w }{3 \left(\bt^2+\tau ^2\right)}\sinh (u) \text{sech}\left(\frac{\bt}{\tau }\right) \text{sech}^3\left(\frac{\bt}{\tau }+u\right).
\end{align}
Fig. \ref{fig:lambda-theta-xi} confirms that we made an appropriate choice \eqref{eq:g1} and \eqref{eq:a1} for $ g_1(\bt) $ and $ a_1(\bt) $. Fig. \ref{fig:gs-fs} shows that $ \bFS $ becomes positive for all times, where the parameters are
\begin{align}\label{eq:parameters}
\tau=10,\quad w=2,\quad u=\frac{1}{10},\quad \epsilon=5.
\end{align}

\begin{figure}[t]
	\centering
	\includegraphics[width=\linewidth]{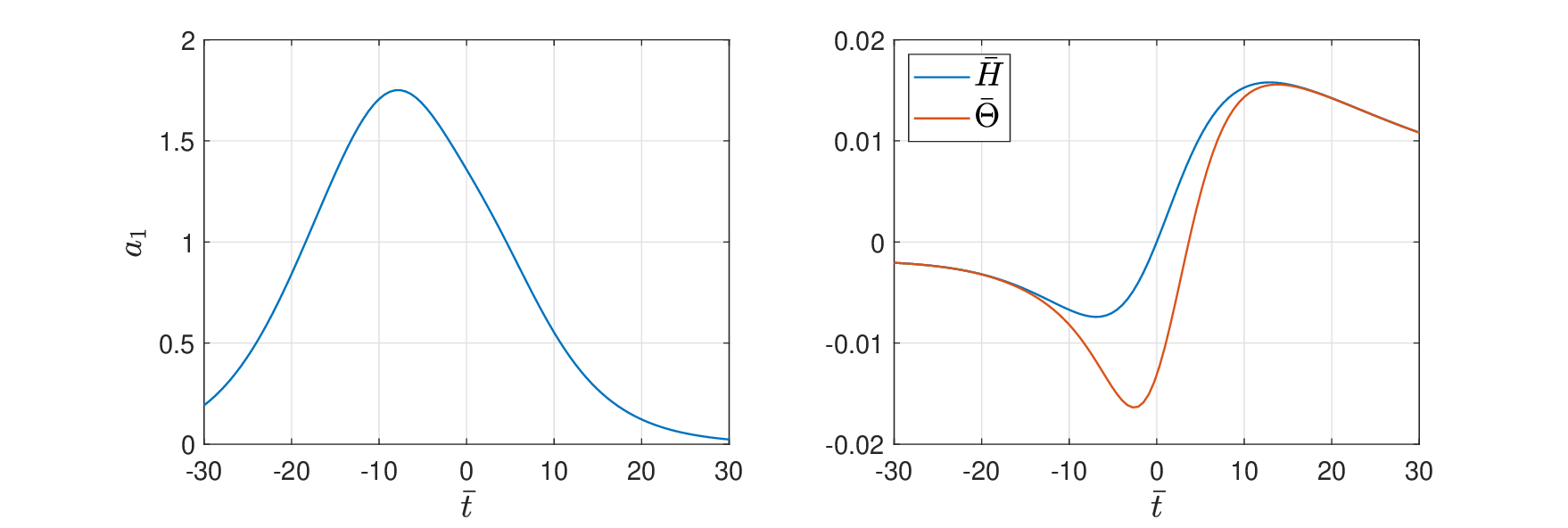}
	\caption{Plot of $ a_1 $, $ \Theta $ and $ \bH $ with the parameters given by \eqref{eq:parameters}.}
	\label{fig:thetatilde}
\end{figure}

\begin{figure}[t]
	\centering
	\includegraphics[width=0.7\linewidth]{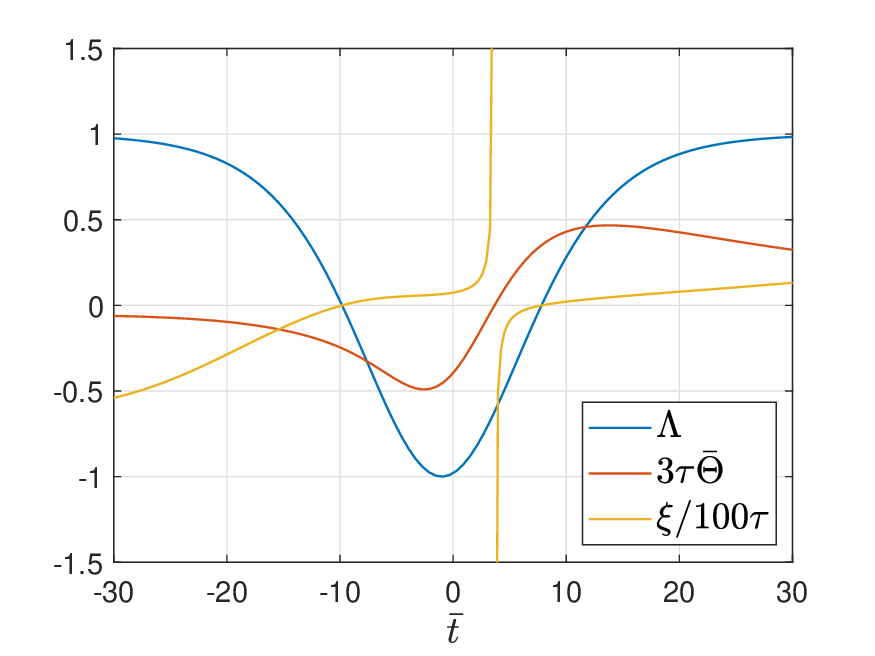}
	\caption{Plot of $ \Lambda $, $ \btTh $ and $ \xi $ with the parameters given by \eqref{eq:parameters}.}
	\label{fig:lambda-theta-xi}
\end{figure}

The next step is to determine the Lagrangian functions $ f_{0,1,2}(\bt) $ such that the constraint $ \bGS\geq \bFS $, which reads
\begin{align}\label{eq:constraint-bGS-bFS}
\btSig\geq \btTh^2\left[\frac{d}{dt}\left(\frac{\Lambda}{\btTh} \right)+\bH\frac{\Lambda}{\btTh}-4\right],
\end{align}
where the function $ \btSig $ given in \eqref{eq:Sigma} is reduced to
\begin{align}\label{eq:btSig}
\btSig=&\; f_1(\bt)+6 f_2(\bt)-2 g_1(\bt) \ddot g_1(\bt)+3 \dot g_1(\bt)^2\nonumber\\
&+ \bH(\bt) \left[(15 a_1(\bt)-4 g_1(\bt)-15) \dot g_1(\bt)-(g_1(\bt)-3) \dot a_1(\bt)\right]\nonumber\\
&-\left[a_1(\bt) (g_1(\bt)-3)+(g_1(\bt)-11) g_1(\bt) \right] \dot\bH(t)\nonumber\\
&-3 \bH^2(\bt) \left[a_1(\bt) (g_1(\bt)+15)+(g_1(\bt)-1) g_1(\bt)+1\right].
\end{align}
The form of $ \btSig $ should be chosen such that not only the constraint \eqref{eq:constraint-bGS-bFS} is satisfied but also it has the correct asymptotics. Once done, \eqref{eq:btSig} will give a relationship between the Lagrangian functions $ f_1(\bt) $ and $ f_2(\bt) $.

It follows from \eqref{eq:asymptotics-H-F} and \eqref{eq:asymptotics-a-g}--\eqref{eq:asymptotics-f2} that $ \btSig $ behaves asymptotically as
\begin{align}\label{eq:asymptotics-tilde-Sigma}
\btSig\sim \frac{\epsilon-3}{\epsilon^2 \bt^2} \text{ for } \bt\to-\infty,\quad 
\Sigma\propto\frac{1}{\bt^4} \text{ for } \bt\to+\infty.
\end{align}
In the limiting case $ |\bt|\to+\infty $, the exponentially suppressed terms can be ignored, and thus the constraint \eqref{eq:constraint-bGS-bFS} becomes
\begin{align}\label{eq:condition-superluminality}
\btSig \geq -\dot \bH-3\bH^2\sim\frac{\bt^2(\gamma-3)-\gamma\tau^2}{\gamma^2(\bt^2+\tau^2)^2},
\end{align}
where $ \gamma=3 $ for $ \bt>0 $ and $ \gamma=\epsilon $ for $ \bt<0 $.
Our choice for $ \btSig $ with the asymptotics \eqref{eq:asymptotics-tilde-Sigma} that satisfies the condition \eqref{eq:condition-superluminality} is
\begin{align}\label{eq:tilde-Sigma-explicit}
\btSig(\bt) = \frac{\tau ^4}{\left(\bt^2+\tau ^2\right)^2}+\frac{(\epsilon -3) \left(1-\tanh \left(\frac{\bt}{\tau }\right)\right)}{2 \epsilon ^2 \left(\bt^2+\tau ^2\right)}.
\end{align}
We numerically checked that $ \bGS>\bFS >0$ and thus $ 0<c_s^2<1 $ for all times, see Fig. \ref{fig:gs-fs} and \ref{fig:cssquared}.

\begin{figure}[t]
	\centering
	\includegraphics[width=0.7\linewidth]{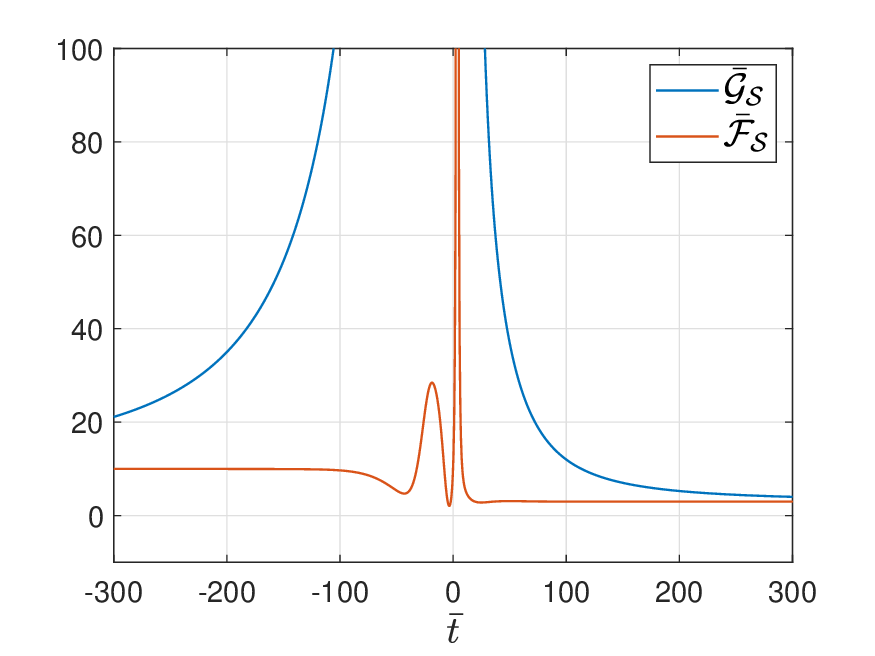}
	\caption{Plot of $ \bGS $ and $ \bFS $. The parameters $ w $, $ u $, $ \tau $ and $ \epsilon $ are given by \eqref{eq:parameters}.}
	\label{fig:gs-fs}
\end{figure}

\begin{figure}[t]
	\centering
	\includegraphics[width=0.7\linewidth]{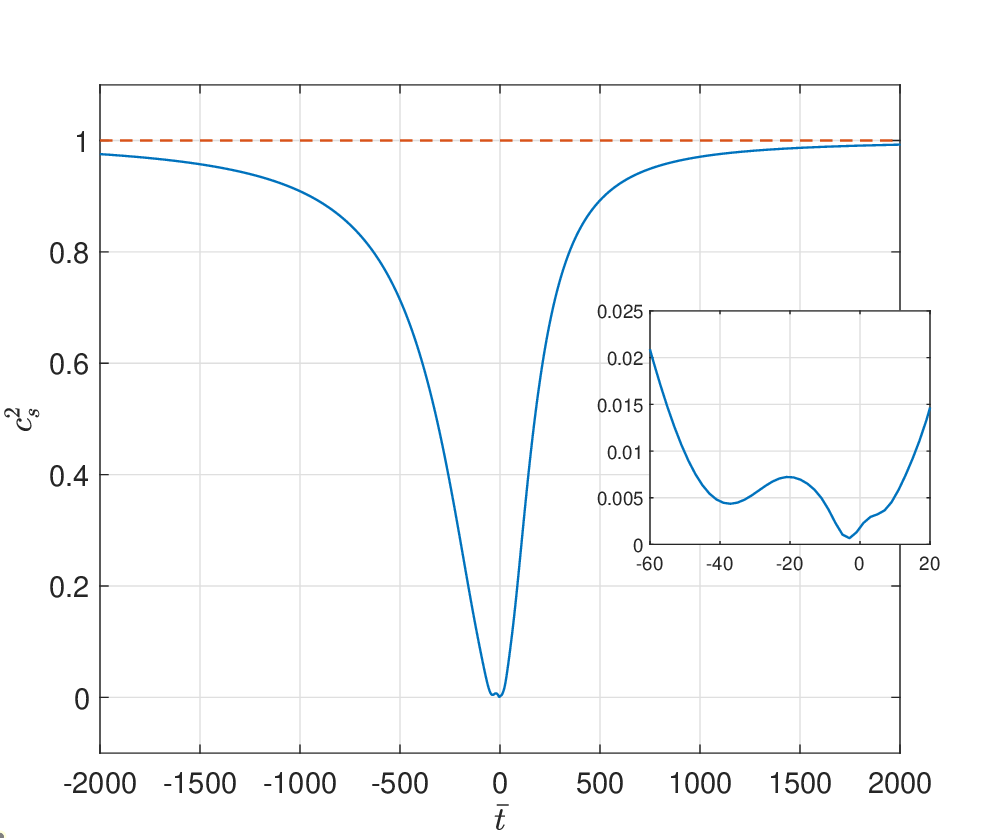}
	\caption{Plot of sound speed squared $ c_s^2 $, which lies between $ 0 $ and $ 1 $ for all times. The main panel shows that $ c_s^2 $ tends to $ 1 $ as $ |\bt|\to+\infty $. The inset shows sound speed squared in the vicinity of bounce at $ \bt=0 $. The parameters $ w $, $ u $, $ \tau $ and $ \epsilon $ are given by \eqref{eq:parameters}.}
	\label{fig:cssquared}
\end{figure}

It follows from \eqref{eq:btSig} and \eqref{eq:tilde-Sigma-explicit} that
\begin{align}\label{eq:f16f2}
f_1(\bt)+6 f_2(\bt)=&\frac{\tau ^4}{\left(\bt^2+\tau ^2\right)^2}+\frac{(\epsilon -3) \left(1-\tanh \left(\frac{\bt}{\tau }\right)\right)}{2 \epsilon ^2 \left(\bt^2+\tau ^2\right)}+2 g_1(\bt) \ddot g_1(\bt)-3 \dot g_1(\bt)^2\nonumber\\
&- \bH(\bt) \left[(15 a_1(\bt)-4 g_1(\bt)-15) \dot g_1(\bt)-(g_1(\bt)-3) \dot a_1(\bt)\right]\nonumber\\
&+\left[a_1(\bt) (g_1(\bt)-3)+(g_1(\bt)-11) g_1(\bt) \right] \dot\bH(t)\nonumber\\
&+3 \bH^2(\bt) \left[a_1(\bt) (g_1(\bt)+15)+(g_1(\bt)-1) g_1(\bt)+1\right].
\end{align}
Here we do not give the explicit form of the Lagrangian functions $ f_{0,1,2} $, which will be determined in the next section, since they are affected by the extra scalar field $ \chi $ through the background equations of motion, though their asymptotic behavior remains unchanged.

\section{Extra Scalar and Comparison with Observations}\label{sec:extra-field}

In this section we consider the coupled case, that is, allow $ \chi $ to have non-zero kinetic energy and interact with $ \phi $ so that the perturbations $ \delta\chi $ can be converted into the curvature fluctuations. 

\subsection{Setup}
As mentioned in the Introduction we focus on the typical case in which $ P(\chi,Y,\phi) $ has the conventional form such that the perturbation $ \delta\chi $ has luminal sound speed, namely
\begin{align}
P(\chi,Y,\phi)=P_1(\chi,\phi)Y+P_0(\chi,\phi),
\end{align}
with $ P_1(\chi,\phi)>0 $. As in the previous section, it is convenient to work on the rescaled quantities:
\begin{align}\label{eq:rescaled-chi}
\chi=M_{pl}\bchi,\quad Y=\mu^4\bY ,\quad P(\chi,Y,\phi)=\mu^4\bP(\bchi,\bY,\bphi)=\mu^4(\bP_0+\bY P_1).
\end{align}
Let us first make an ansatz for $ \bP_0(\bchi,\bphi) $ as follows:
\begin{align}
\bP_0(\bchi,\bphi)=-\bP_{02}(\bphi)\bchi^2-\bP_{04}(\bphi)\bchi^4.
\end{align}
The presence of the quartic term is only for stability of the scalar potential. Of course, it is possible to make the scalar potential bounded from below only in terms of the quadratic term. It, however, turns out to be highly non-trivial in this case to make the conversion from $ \delta\chi $ into $ \tilde\zeta $ sufficiently efficient. Note that in order for interaction energy $ -P_0(\bchi,\bphi) $ to be bounded from below, $ \bP_{04}(\bphi) $ should be positive and $ -\bP_{02}^2(\bphi)/\bP_{04}(\bphi) $ should be bounded from below. We also assume that $ P_1 $ does not depend on $ \chi $ and the scalar field $ \chi $ has canonical kinetic term in the distant future. As we see below, the behavior of $ P_1 $ in the distant past is not canonical.

The background equations of motion \eqref{eq:background-g00}, \eqref{eq:background-gii} and \eqref{eq:background-chi} become
\begin{subequations}\label{eq:background-simple-two-fields}
	\begin{align}
	& \left(f_0(\bt)-\bP_{02}(\bt)\bchi^2-\bP_{04}(\bt)\bchi^4\right)-\left(f_1(\bt)+\bY P_1(\bt)\right)-3 f_2(\bt)\nonumber\\
	&\hskip10em+3\bH\btTh+3\bH\dot g_1-3g_1(2\dot \bH+3\bH^2)=0,\label{eq:background-bar-g00}\\
	& \left(f_0(\bt)-\bP_{02}(\bt)\bchi^2-\bP_{04}(\bt)\bchi^4\right)+\left(f_1(\bt)+\bY P_1(\bt)\right)+f_2(\bt)+2 \dot \bH+3 \bH^2=0,\label{eq:background-bar-gii}\\
	& \frac{1}{a^3}\frac{d}{d\bt}\left(a^3\dot\bchi P_1(\bt)\right)+\bP_{02}(\bt)\bchi+2\bP_{04}(\bt)\bchi^3=0.\label{eq:background-bchi}
	\end{align}
\end{subequations}
For simplicity, we let $ \dot\chi $ decay exponentially in the asymptotic limit so that $ \tilde\zeta $ and $ \delta\chi $ decouple both in the distant past and in the distant future. It then follows from \eqref{eq:background-bchi} that $ \bP_{02}(\bt) $ and $ \bP_{04}(\bt) $ are also suppressed exponentially. This offers two advantages. The first one is that the asymptotic behavior of $ f_{0,1,2}(\bt) $ derived in the previous subsection is not affected through the background equations of motion \eqref{eq:background-bar-g00} and \eqref{eq:background-bar-gii}. The second one is that it is guaranteed that $-\bP_{02}^2(\bphi)/\bP_{04}(\bphi) $ is bounded from below as long as $ \bP_{04} $ does not touch zero.

Our goal is to choose $ \bchi(\bt) $ and the Lagrangian functions $ P_1 $, $ \bP_{02} $ and $ \bP_{04}>0 $ such that
\begin{itemize}
	\item a nearly scale-invariant power spectrum of the extra scalar field $ \chi $ is generated during the ekpyrotic phase,
	\item the perturbations $ \delta\chi $ are converted into the curvature fluctuations $ \tilde\zeta $ with high efficiency by allowing $ \chi $ to have non-zero kinetic energy and interact with $ \bphi $,
	\item the background equations of motion \eqref{eq:background-simple-two-fields} and the constraint \eqref{eq:positivity-FAB} are satisfied,
	\item $ \dot\bchi(\bt) $, $ \bP_{02}(\bt) $ and $ \bP_{04}(\bt) $ decay exponentially with the same exponent as $ |\bt|\to+\infty $.
\end{itemize}

\subsection{Scalar spectral index for entropic perturbations}

First of all, the nearly scale-invariant perturbations can be generated by taking the form of $ P_1(\chi,\phi) $ as $ \phi^{-2(1+b)} $ with $ b\ll 1 $  \cite{Qiu:2013eoa,Li:2013hga,Fertig:2013yyy,Fertig:2016czu}. To be concrete, in the decoupled case the equation for the Fourier mode $ \delta\chi_k $ of the perturbation $ \delta\chi $ becomes
\begin{align}
\frac{1}{a^3}\frac{d}{dt}\left(a^3 P_1 \frac{d}{dt}\delta\chi_k\right)+\frac{k^2}{a^2}P_1 \delta\chi_k=0,
\end{align}
or
\begin{align}\label{eq:vk-ekpyrotic}
\frac{d^2}{d\eta^2}v_k+\left(k^2-\frac{1}{z}\frac{d^2 z}{d\eta^2}\right)v_k=0,
\end{align}
where $ v_k\equiv z\delta\chi_k $ with $ z\equiv a\sqrt{P_1} $ and $ \eta $ denotes the conformal time defined by $ dt=a\;d\eta $. Note that during the ekpyrotic contraction phase, the conformal time $ \eta $ is related to $ t $ by
\begin{align}
-\eta\sim (-t)^{\frac{\epsilon-1}{\epsilon}}.
\end{align}
It follows from $ P_1\sim (-\bt)^{-2(1+b)} $ that $ z\sim (-\eta)^{-b\epsilon/(\epsilon-1)-1} $. Since in the distant past the mode functions obey the equation for a free field in Minkowski space, we can impose the Bunch-Davies vacuum initial condition for the mode functions, namely
\begin{align}
\lim_{k\eta\to-\infty}v_k=\frac{1}{\sqrt{2k}}e^{-ik\eta}.
\end{align}
The solution of \eqref{eq:vk-ekpyrotic} is then
\begin{align}
v_k=\frac{\sqrt{-\pi\eta}}{2}H_\nu^{(1)}(-k\eta)\quad \text{with } \nu=\frac32+\frac{b\epsilon}{\epsilon-1},
\end{align}
where $ H_\nu^{(1)} $ is a Hankel function of the first kind. In the super-horizon limit, i.e. $ |k\eta|\ll 1 $, we obtain
\begin{align}
v_k\sim \frac{(-\eta)^{\frac12-\nu}}{k^\nu},
\end{align}
and thus we find the spectral index $ n_s $ related to $ b $ by
\begin{align}\label{eq:ns-b}
n_s=1-\frac{2b\epsilon}{\epsilon-1}.
\end{align}
Since $ \epsilon\gg 1 $, for the value of $ b $ around $ 0.0158 $ the spectral tilt becomes $ n_s\simeq 0.965 $, which is in good agreement with the Planck data \cite{Aghanim:2018eyx}. Eventually, once $ P_1(\bphi)=P_1(\bt) $ behaves in the far past as
\begin{align}
P_1(\bt)\propto \frac{1}{(-\bt)^{2(1+b)}}\quad \text{ for } \bt\to-\infty,
\end{align}
the power spectrum of $ \delta\bchi $ is nearly scale-invariant. 

\subsection{Conversion of entropic perturbations into curvature fluctuations}
Note that during the ekpyrotic phase the curvature fluctuation $ \tilde\zeta $ is highly blue-shifted, namely
\begin{align}
a\sqrt{\GS}M_{pl}\tilde\zeta_k=\frac{\sqrt{-\pi\eta}}{2}H_{\nu_1}^{(1)}(-k\eta)\quad \text{with}\quad \nu_1=\frac{\epsilon-3}{2(\epsilon-1)},\label{eq:nu-1}
\end{align}
where $ 0<\nu_1<\frac12 $, given $ \epsilon>3 $. It follows that $ a\sqrt{\GS}M_{pl}\tilde\zeta_k\sim \frac{(-\eta)^{\frac12-\nu_1}}{k^{\nu_1}} $ for $ |k\eta|\ll 1 $, and hence the amplitude of $ \tilde\zeta $ is much smaller than the one of $ \delta\bchi $ during the ekpyrotic phase. We therefore need to introduce a mechanism to convert $ \delta\bchi $ into $ \tilde\zeta $ with high efficiency.

We choose the following form of $ P_1(\bt) $ (see Fig. \ref{fig:p1})
\begin{align}\label{eq:P1}
P_1(\bt) = &\;\frac12\frac{\sigma \left(-\frac{\bt}{\tau }\right)}{ \left(\frac{\bt}{\tau }\right)^{2(1+b)}+1}+\frac{1}{2}\sigma \left(\frac{\bt}{\tau }\right),
\end{align}
so that the scalar field $ \chi $ has canonical kinetic energy in the far future.

\begin{figure}
	\centering
	\includegraphics[width=0.7\linewidth]{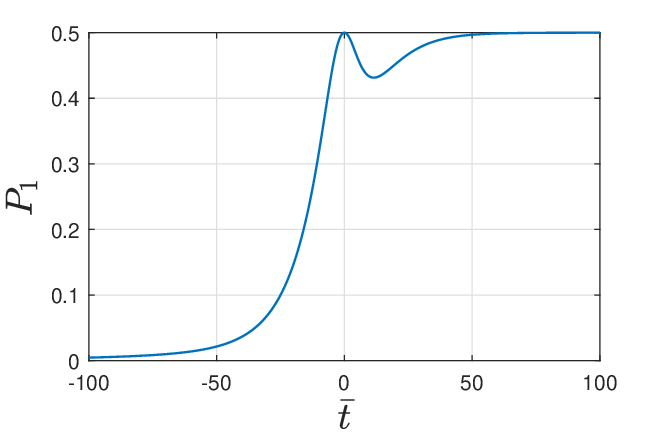}
	\caption{Plot of $P_1$. The parameters are set to be $ \tau=10 $ and $ b=0.0158 $, respectively.}
	\label{fig:p1}
\end{figure}

Now we need to choose $ \bchi $, $ \bP_{02} $ and $ \bP_{04} $ such that conversion from the perturbation $ \delta\bchi $ into the curvature fluctuation $ \tilde\zeta $ occurs efficiently. We remind that $ \bP_{04} $ is introduced solely for stability of the potential and $ \bP_{02} $ is determined by the background equation of motion \eqref{eq:background-bchi} once $ \bchi $ and $ \bP_{04} $ are specified. It is therefore crucial to choose the form of $ \bchi $ appropriately.

Let us first investigate what is the main factor in conversion from entropy to curvature perturbations. It follows from \eqref{eq:quadratic-action-perturbations} that the scalar perturbations obey the following equations:
\begin{subequations}\label{eq:perturbations-full}
	\begin{align}
	\frac{1}{a^3}\frac{d}{d\bt}\left[a^3\left(\left(\bGS+\frac{\bY P_1}{\btTh^2}\right)\tdzetak+\dot\bchi P_1 \bg\dbdchik+\frac12 \bPsi_1\dbchik \right)\right]+\hskip3em&\nonumber\\
	+\frac{\bk^2}{a^2}\left(\bFS\tzetak+\dot\bchi P_1 \bg \dbchik \right)\;&=0,\\
	\frac{1}{a^3}\frac{d}{d\bt}\left[a^3\left(P_1\dbdchik+\dot\bchi P_1 \bg\tdzetak\right) \right]+\frac{\bk^2}{a^2}\left(P_1\dbchik +\dot\bchi P_1 \bg\tzetak\right)-\frac12 \bPsi_1\tdzetak-\bPsi_2\dbchik\;&=0,
	\end{align}
	where $ \bk=\frac{M_{pl}k}{\mu^2 } $ is the (dimensionless) rescaled wavenumber and
	\begin{align}
	\bPsi_1=&\;\frac{2}{\btTh^2}\left[(\bY P_1+\btSig)\dot\bchi P_1-\btTh (\bP_{02}\bchi+2\bP_{04}\bchi^3) (1+3\Delta) \right] ,\label{eq:bPsi1}\\
	\bPsi_2=&\;-\bP_{02}-6\bP_{04}\bchi^2+\frac{1}{a^3}\frac{d}{d\bt}\left(\frac{a^3 \bY P_1^2\left(1-3\Delta\right)}{\btTh}\right)\nonumber\\
	&\hskip1em+\frac{\dot\bchi P_1}{\btTh^2}\left[\dot\bchi P_1(\bY P_1+\btSig)+\btTh P_{0\bchi}(1+3\Delta) \right],\\
	\bg = &\;-\frac{1}{\btTh}\left(1-3\Delta\right).
	\end{align}
\end{subequations}

In the long wavelength limit $ \bk\to0 $, which is of observational interest, the equations \eqref{eq:perturbations-full} become
\begin{subequations}
	\begin{align}
	& \left(\bGS+\frac{\bY P_1}{\btTh^2}\right)\tdzetak+\dot\bchi P_1 \bg\dbdchik+\frac12 \bPsi_1\dbchik=\frac{C_0}{a^3},\\
	& \frac{1}{a^3}\frac{d}{d\bt}\left[a^3\left(P_1\dbdchik+\dot\bchi P_1 \bg\tdzetak\right) \right]-\frac12 \bPsi_1\tdzetak-\bPsi_2\dbchik=0,
	\end{align}
\end{subequations}
or
\ifx
\begin{align}
& a^3\left(\bG_{11}\dot{\tilde\zeta}+\bG_{12}\dot{\delta\bchi}+\frac12\bPsi_1\delta\bchi\right)=C_0,\\
& \left(P_1-\frac{\bG_{12}^2}{\bG_{11}}\right)\ddot{\delta\bchi}+\left(\dot P_1-\frac{d}{d\bt}\left(\frac{\bG_{12}^2}{\bG_{11}}\right)\right)\dot{\delta\bchi}-\frac12 \frac{d}{d\bt}\left(\frac{\bG_{12}\bPsi_1}{\bG_{11}}\right)\delta\bchi-\frac{\bG_{12}\bPsi_1}{2\bG_{11}}\dot{\delta\bchi}\nonumber\\
&\hskip3em+\frac{C}{a^3}\frac{d}{d\bt}\left(\frac{\bG_{12}}{\bG_{11}}\right)+3H\left[\left(P_1-\frac{\bG_{12}^2}{\bG_{11}}\right)\dot{\delta\bchi}-\frac{\bG_{12}}{2\bG_{11}}\bPsi_1\delta\bchi\right]\nonumber\\
&\hskip3em-\frac12\frac{\bPsi_1}{\bG_{11}}\left(\frac{C}{a^3}-\bG_{12}\dot{\delta\bchi}-\frac{1}{2}\bPsi_1\delta\bchi\right)-\bPsi_2\delta\bchi=0
\end{align}
\fi
\begin{subequations}\label{eq:perturbations-simple}
	\begin{align}
	& 0= \tdzetak-\frac{1}{\bG_{11}}\left(\frac{C_0}{a^3}-\bG_{12}\dbdchik-\frac12 \bPsi_1\dbchik\right),\label{eq:tildezeta-k}\\
	&0= \left(P_1-\frac{\bG_{12}^2}{\bG_{11}}\right)\ddot{\delta\bchi}_k+\left[3\bH\left(P_1-\frac{\bG_{12}^2}{\bG_{11}}\right)+\dot P_1-\frac{d}{d\bt}\left(\frac{\bG_{12}^2}{\bG_{11}}\right)\right]\dbdchik\nonumber\\
	&\hskip3em+\left[\frac{\bPsi_1^2}{4\bG_{11}}-3\bH\frac{\bG_{12}}{2\bG_{11}}\bPsi_1-\frac12 \frac{d}{d\bt}\left(\frac{\bG_{12}\bPsi_1}{\bG_{11}}\right)-\bPsi_2\right]\dbchik\nonumber\\
	&\hskip3em+\frac{C_0}{a^3}\left[\frac{d}{d\bt}\left(\frac{\bG_{12}}{\bG_{11}}\right)-\frac12\frac{\bPsi_1}{\bG_{11}}\right],
	\end{align}
\end{subequations}
where $ C_0 $ is an integration constant, which is determined by the initial condition, and
\begin{align}
\bG_{11}= \bGS+\frac{\bY P_1}{\btTh^2},\quad \bG_{12}=\dot\bchi P_1 \bg.
\end{align}

It is now evident from \eqref{eq:tildezeta-k} that $ \bPsi_1/\bG_{11} $ is the main factor that controls the efficiency of conversion process. Since $ \bPsi_1 $ depends on $ \dot{\bchi} $, one is tempted to make $ |\dot{\bchi}| $ large. This approach has drawbacks, though, due to the constraint \eqref{eq:positivity-FAB}. Therefore, the only possible way to increase $ \bPsi_1 $ is to make $ |\bP_{02}\bchi+2\bP_{04}\bchi^3|=|\frac{d}{d\bt}(a\dot{\bchi}P_1)| $ large by increasing $ \ddot{\bchi} $ instead of $ \dot{\bchi} $. For this reason, we choose the following form of $ \bchi $ (see Fig. \ref{fig:chi})
\begin{align}\label{eq:chi}
\bchi=\bchi_0+\lambda\, \text{sech}\left(\frac{\bt-\bt_c}{2 \tau }\right)\left[e^{-\left(\frac{\bt-\bt_c}{4 \tau }\right)^2}-q\right],
\end{align}
where the parameter $ \lambda $ controls the overall rate of change of $ \bchi $ and the parameter $ \bt_c $ is related to the moment of conversion from $ \delta\chi $ into $ \tilde\zeta $. We set $ \bt_c\sim 10\tau $ so that the conversion occurs sufficiently long after the bounce phase. The bias $ \bchi_0 $ is introduced in order for $ \bchi $ not to cross zero; otherwise $ \bP_{02} $ and $ \bP_{04} $ become singular at a certain point. The overall shape of the background trajectory around the conversion phase is determined by the function $ \text{sech}\left(\frac{\bt-\bt_c}{2 \tau }\right) $ in \eqref{eq:chi}. On the other hand, the factor in the square bracket plays two roles, the first one as enhancing the conversion efficiency and the second one as making $ \bP_{02} $ positive in the asymptotic limit. The second role, which is actually played by the parameter $ q $, is necessary since we do not want the tachyonic instability to appear in the asymptotic part of the background trajectory of interest. By adjusting the parameters $ \bchi_0 $, $ \lambda $ and $ q $, we can make the conversion efficient and smooth, while not violating the constraint \eqref{eq:positivity-FAB}. Here we mean the conversion efficiency by \cite{Fertig:2015ola,Fertig:2016czu}
\begin{align}\label{eq:def-conv-efficiency}
|\tzetak(\bt_{conv-end})/\dbchik(\bt_{conv-begin})|,
\end{align}
where $ \bt_{conv-begin}  $ and $ \bt_{conv-end} $ denote the beginning and end moment of the conversion process, respectively. And the conversion is said to be smooth when the comoving Hubble radius $ (aH)^{-1} $ changes by order of one e-fold during conversion \cite{Lehners:2007wc,Fertig:2015ola}.

\begin{figure}[t]
	\centering
	\includegraphics[width=\linewidth]{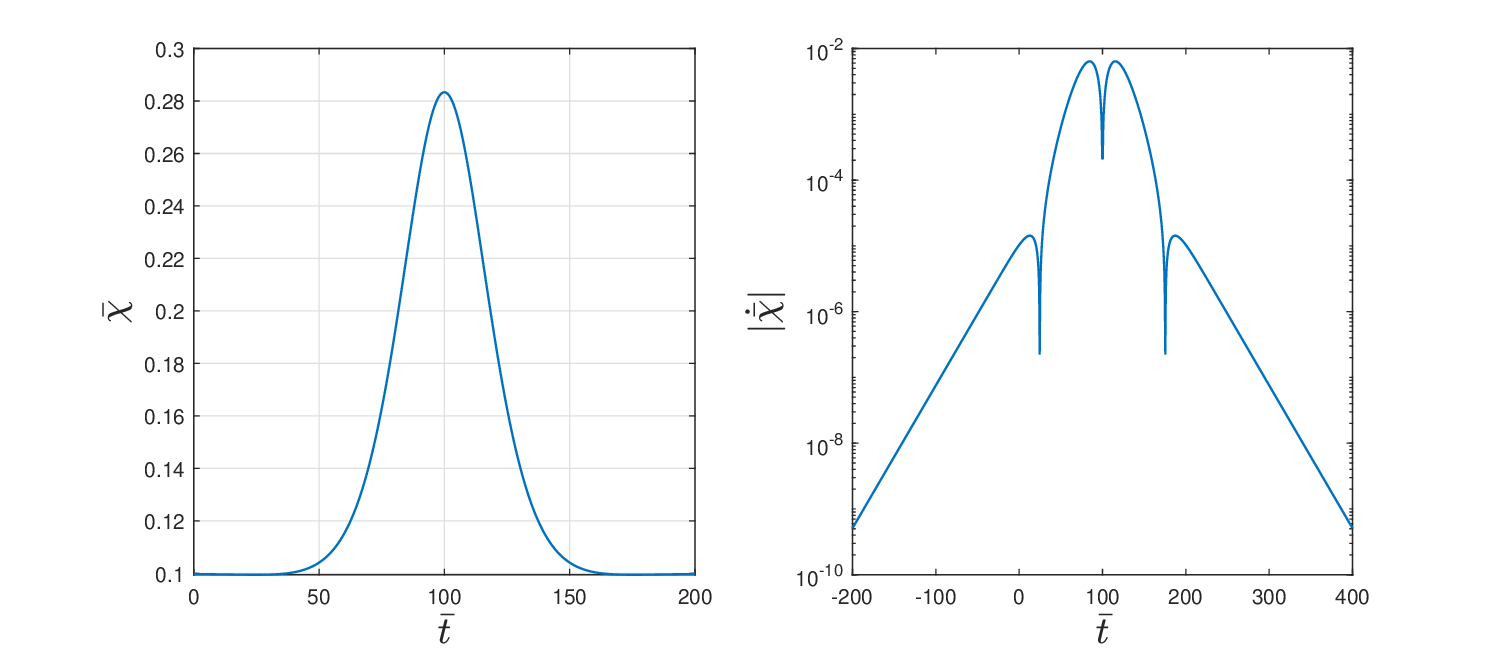}
	\caption{Plot of $\bchi$ and $ \dot{\bchi} $. The parameters $ \bchi_0 $, $ \lambda $, $ \tau $ and $ \bt_c $ are given by \eqref{eq:parameters-full}.}
	\label{fig:chi}
\end{figure}

We choose the following form of $ \bP_{04} $
\begin{align}\label{eq:P04}
\bP_{04}(\bt) = 10^{-10}\text{sech}\left(\frac{\bt-\bt_c}{2\tau}\right),
\end{align}
so that it has the same asymptotics with $ \dot{\bchi} $ and plays little role in the dynamics. The Lagrangian function $ \bP_{02} $ is determined by combining \eqref{eq:background-bchi}, \eqref{eq:chi} and \eqref{eq:P04}, namely
\begin{align}\label{eq:P02}
\bP_{02}(\bt)=-2\bP_{04}(\bt)\bchi^2-\frac{1}{a^3 \bchi}\frac{d}{d\bt}\left(a^3\dot\bchi P_1(\bt)\right).
\end{align}
One can readily see that the asymptotic behavior of $ \bP_{02} $ and $ \bP_{04} $ is given as follows:
\begin{align}\label{eq:asymptotics-P02-P04}
& \bP_{02}\sim \frac{1}{\bt^2}\exp\left(\frac{\bt}{2\tau}\right) \text{ as } \bt\to-\infty,\quad \bP_{02}\sim \exp\left(\frac{-\bt}{2\tau}\right) \text{ as } \bt\to+\infty,\\
& \bP_{04}\sim \exp\left(-\frac{|\bt|}{2\tau}\right) \text{ as } |\bt|\to+\infty.
\end{align}
Fig. \ref{fig:p02} confirms that $ \bP_{02} $ and $ \bP_{04} $ behave asymptotically as given in \eqref{eq:asymptotics-P02-P04}. It follows from \eqref{eq:asymptotics-P02-P04} that $ -\bP_{02}^2/\bP_{04} $ is bounded from below, see Fig. \ref{fig:p02squared-p04}.

\begin{figure}[t]
	\centering
	\includegraphics[width=\linewidth]{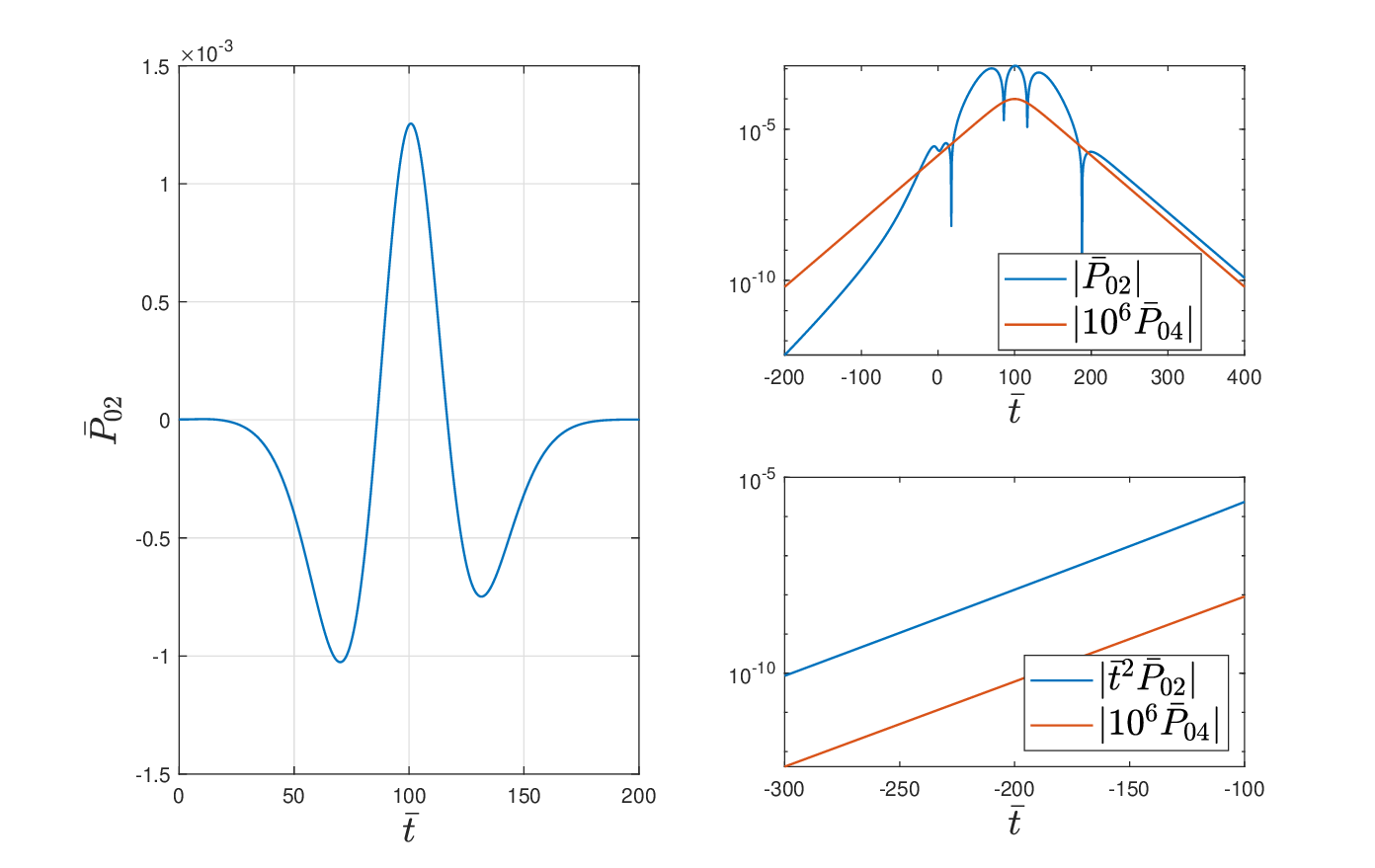}
	\caption{The left panel shows plot of $ \bP_{02} $, while the right two panels show $ \bP_{02} $ and $ \bP_{04} $ behave asymptotically as given in \eqref{eq:asymptotics-P02-P04}. The parameters are given by \eqref{eq:parameters-full}.}
	\label{fig:p02}
\end{figure}

\begin{figure}[t]
	\centering
	\includegraphics[width=0.7\linewidth]{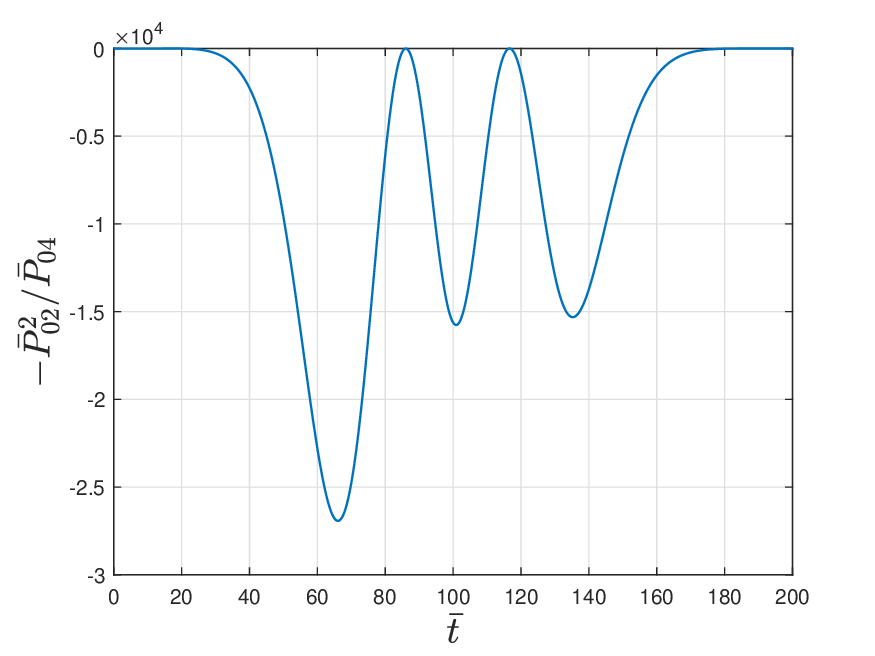}
	\caption{Plot of $-\bP_{02}^2/\bP_{04}$. The parameters are given by \eqref{eq:parameters-full}.}
	\label{fig:p02squared-p04}
\end{figure}

Now that all Lagrangian functions and the background evolution of the fields are specified, we solve the equations \eqref{eq:perturbations-simple} numerically to confirm that $ \delta\chi $ is converted into $ \tilde\zeta $ with high efficiency. The initial condition is
\begin{align}
\tilde\zeta_k(\bt_i)=10^{-5}\sqrt{\frac{P_1(\bt_i)}{\bGS(\bt_i)}},\quad \delta\bchi_k(\bt_i)=10^{-5},\quad \dot{\tilde\zeta}_k(\bar t_i)=0,\quad \dbdchik(\bt_i)=0,
\end{align}
where $ \bt_i $ is the initial time. The parameters are set to be \begin{subequations}\label{eq:parameters-full}
	\begin{align}
	& \tau=10,\quad \epsilon=10,\quad u=\frac{1}{10},\quad w=2,\\
	& b=0.0158,\quad \bt_c=10\tau=100,\quad q=\frac{1}{12},\quad  \lambda=0.35,\quad \bchi_0=\frac{1}{10},
	\end{align}
\end{subequations}
where the values of $ \tau $, $ u $, and $ w $ are the same as those given in \cite{Mironov:2018oec}, while the values of parameters $ \bt_c $, $ q $, $ \lambda $ and $ \bchi_0 $ are chosen based on an educated guess obtained by conducting some tentative numerical analysis. Fig. \ref{fig:perturbations} shows the result of the numeric computation, where the initial time is $\bt_i=-3\times 10^3$. The conversion process takes place around $ 4\tau=\bt_c-6\tau\lesssim \bt\lesssim\bt_c+6\tau=16\tau $, during which about $ 98\% $ of $ \dbchik $ is converted into $ \tzetak $ and the comoving Hubble radius $ (aH)^{-1} $ decreases by order of one e-fold. This implies that the conversion is smooth and highly efficient, cf. \cite{Fertig:2016czu}.

\begin{figure}
	\centering
	\includegraphics[width=\linewidth]{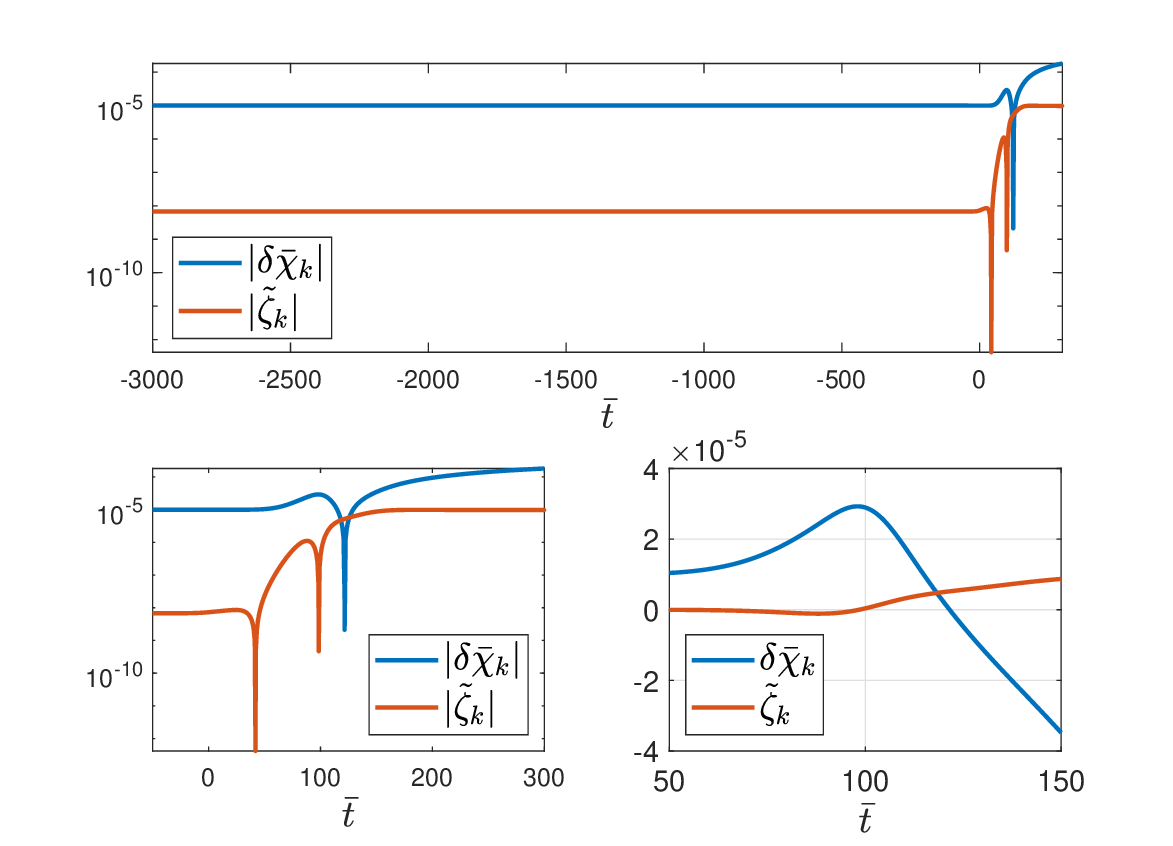}
	\caption{Plot of evolution of $ \tzetak $ and $ \dbchik $. We set the initial time $ \bt_i=-3\times 10^3 $. The parameters are given by \eqref{eq:parameters-full}.}
	\label{fig:perturbations}
\end{figure}

Fig. \ref{fig:positivity-fab} shows that for the parameters given by \eqref{eq:parameters-full} the constraint \eqref{eq:positivity-FAB} is satisfied.

\begin{figure}[t]
	\centering
	\includegraphics[width=0.9\linewidth]{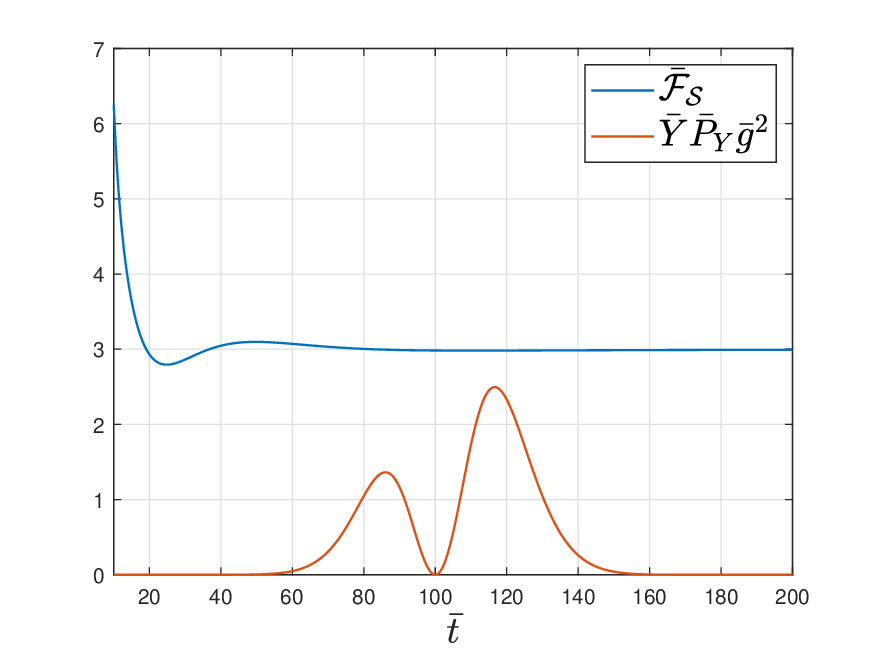}
	\caption{Plot of $\bFS$ and $\bY \bP_Y\bar g^2 $. The function $\bFS$ always dominates over $\bY \bP_Y\bar g^2 $ throughout the whole time evolution, which shows the constraint \eqref{eq:positivity-FAB} is satisfied. The parameters are given by \eqref{eq:parameters-full}.\label{fig:positivity-fab}}
\end{figure}

The remaining functions $ f_{0,1,2}(\bt) $ are uniquely determined from \eqref{eq:btSig} and the background equations of motion \eqref{eq:background-bar-g00} and \eqref{eq:background-bar-gii}. We emphasize that non-trivial form of $ \bchi $, $ \bP_{02} $ and $ \bP_{04} $ does not alter the asymptotics of $ f_{0,1,2}(\bt) $ given in \eqref{eq:asymptotics-f2}, \eqref{eq:asymptotics-f0-f1-past} and \eqref{eq:asymptotisc-f0-f1-future}.

\subsection{The amplitude of scalar power spectrum and characteristic energy scale}
Now we determine the characteristic energy scale $ \mu $ which is related to the amplitude of the perturbations $ \delta\bchi $. Since $ \delta\bchi $ is almost completely converted into $ \tilde\zeta $, the power spectrum of $ \delta\bchi $ should be
\begin{align}
\Delta_{\delta\chi}^2(k)\equiv k^3\braket{\delta\bchi_k^2}=A_s\left(\frac{k}{k_\star}\right)^{n_s-1},
\end{align}
where $ k_\star$ denotes the pivot scale $0.05\text{Mpc}^{-1}\simeq 10^{-59}M_{pl} $, and $ A_s $ denotes the measured amplitude at $ k_\star $, namely \cite{Aghanim:2018eyx}
\begin{align}\label{eq:AS-planck}
A_s=(2.105\pm0.030)\times 10^{-9}.
\end{align}
After the mode $ v_k $ exits the horizon, its amplitude can be written as
\begin{align}
a\sqrt{P_1}M_{pl}\delta\bchi_k=v_k\simeq \frac{\sqrt{-\pi\eta}}{2}\frac{1}{(-k\eta)^\nu}.
\end{align}
On the other hand, the conformal time $ \eta $ is related to the rescaled cosmic time $ \bt $ by
\begin{align}
\left(\frac{-\bt}{\tau}\right)^{1-\frac 1\epsilon}\sim \frac{\epsilon-1}{\epsilon}\frac{\mu^2}{M_{pl}}\left(\frac{-\eta}{\tau}\right).
\end{align}
It then follows that
\begin{align}
a\sqrt{P_1}\sim\left[\frac{\epsilon-1}{\epsilon}\frac{\mu^2}{M_{pl}}\left(\frac{-\eta}{\tau}\right)\right]^{\frac12-\nu},
\end{align}
and thus
\begin{align}\label{eq:power-spectrum-bchi}
\Delta_{\delta\chi}^2(k)\sim\frac{\pi}{4}\left(\frac{\epsilon-1}{\epsilon\tau}\right)^2\left(\frac{\mu}{M_{pl}}\right)^4\left(\frac{\epsilon-1}{\epsilon\tau}\frac{\mu^2}{M_{pl}k_\star}\right)^{1-n_s}\left(\frac{k}{k_\star}\right)^{n_s-1}.
\end{align}
It follows from \eqref{eq:power-spectrum-bchi} that for $ \mu\sim\cO(10^{-2})M_{pl} $ the amplitude of the scalar power spectrum is in good agreement with the Planck data \eqref{eq:AS-planck}.

For completeness, we discuss the tensor-to-scalar ratio. Since $ \bGT=\bFT=1 $ in our scenario, it is evident that the tensor perturbations carry a deep blue spectral index (actually equal to $ 4-2\nu_1$, where $ \nu_1 $ is given by \eqref{eq:nu-1}). It then follows that at the pivot scale the tensor-to-scalar ratio is negligibly small, which is in fact a typical feature of bouncing scenarios that begin with the ekpyrotic contraction phase. Our scenario is therefore compatible with the observational data, regarding the tensor-to-scalar ratio as well.

\section{Summary and Conclusion}\label{sec:conclusion}

In this paper we have constructed a class of DHOST theories with an extra scalar field, which admits viable bouncing solutions over which linear perturbations are free of all pathologies during entire evolution and compatible with observations. In particular, after finding the sufficient conditions for the scalar sector to be stable and non-superluminal (see \eqref{eq:stability-condition-scalar-final}), we constructed for the first time a completely healthy two-field DHOST bounce model, in which the adiabatic field $ \phi $ is described by the DHOST theories and the entropic field $ \chi $ has luminal sound speed. Our model exhibits the conventional asymptotics for the DHOST sector without requiring strong gravity in the past employed in \cite{Ageeva:2018lko,Ageeva:2020buc,Ageeva:2020gti,Ageeva:2021yik,Ageeva:2022fyq,Ageeva:2022asq}. The overview of structure of the model is given in Table \ref{modelstructure}.

\begin{center}
	\begin{spectblr}[caption = {Structure of the two-field model constructed, see \eqref{eq:action}, \eqref{eq:lagrangian-functions} and \eqref{eq:lagrangian-functions-lower}. \label{modelstructure}}]{colspec={c|[2pt,white]c|[2pt,white]c|[2pt,white]c|[2pt,white]c|[2pt,white]c|[2pt,white]c|[2pt,white]c|[2pt,white]c|[2pt,white]c}}
		\SetCell[c=10]{c} The exceptional subclass of DHOST Ia $ + $ an extra field & & & & & & & & & \\
		\SetRow{gray8} \SetCell[c=7]{c} DHOST sector & & & & & & & \SetCell[c=3]{c} Extra field sector & & \\
		\SetRow{azure8} \SetCell[c=3]{c} $ F(\phi,X) $ & & &\SetCell[c=2]{c} $ F_2(\phi,X) $ & & \SetCell[c=2]{c} $ A_1(\phi,X) $ & & \SetCell[c=3]{c} $ P(\chi,Y,\phi) $ & &  \\
		\SetRow{olive9} $ f_0(\phi) $ & $ f_1(\phi) $ & $ f_2(\phi) $ & $ g_0(\phi) $ & $ g_1(\phi) $  & $ a_0(\phi) $ & $ a_1(\phi) $ & $ P_{02}(\phi) $  & $ P_{04}(\phi) $  & $ P_1(\phi) $ 
	\end{spectblr}
\end{center}

As the main calculations in the previous sections might look quite technical, in the flowchart in Fig. \ref{fig:flowchart} we summarized the essentials/guideline of  how the model has been constructed using the inverse method.
\begin{figure}[t]
	\centering
	\begin{tikzpicture}[node distance=5mm,>={Stealth[round]},
	terminalDHOST/.style={
		rectangle,minimum width=3cm, text width=2.5cm, rounded corners,drop shadow,
		very thick,draw=teal!70!black,
		top color=white,bottom color=teal!60,
		font=\small, align=center},
	terminalCHI/.style={
		rectangle,minimum width=3cm, text width=2.5cm, rounded corners,drop shadow,
		very thick,draw=orange!60!black,
		top color=white,bottom color=orange!60!yellow,
		font=\small, align=center},
	requirements/.style={
		rectangle,minimum width=9.5cm, text width=9.5cm, rounded corners, drop shadow,
		very thick,draw=cyan!70!black,
		top color=white,bottom color=cyan!40,
		font=\small, align=center}]
	\node (phi) [terminalDHOST] {$ \phi $: clock\\\eqref{eq:phi}};
	\node (a) [terminalDHOST, below=of phi] {Scale factor $ a $\\\eqref{eq:scale-factor}};
	\node (a0g0) [terminalDHOST, below=of a] {$ g_0=-g_1 $, $ a_0=-a_1 $\\\eqref{eq:g0-a0}};
	\node (a1g1) [terminalDHOST, below=of a0g0] {$ g_1 $, $ a_1 $\\\eqref{eq:g1}, \eqref{eq:a1}};
	\node (Sigma) [terminalDHOST, below=of a1g1] {$ f_1+6f_2 $\\\eqref{eq:f16f2}};
	\node (P1) [terminalCHI, below=of Sigma] {$ P_1$\\\eqref{eq:P1}};
	\node (chi) [terminalCHI, below=of P1] {$ \bchi $\\\eqref{eq:chi}};
	\node (P04) [terminalCHI, below=of chi] {$ \bP_{04} $\\\eqref{eq:P04}};
	\node (P02f012) [terminalCHI, below=of P04] {$ \bP_{02}, f_0, f_1, f_2 $\\\eqref{eq:P02-explicit}--\eqref{eq:f0-explicit}};
	\node (BKL) [requirements, right=2cm of a] {Bounce solution with ekpyrotic contraction leading to the absence of BKL instability};
	\node (tensor) [requirements, right=2cm of a0g0] {Stability and non-superluminality in tensor sector:\\$ \bGT=\bFT=1 $};
	\node (DHOSTscalarInstability) [requirements, right=2cm of a1g1] {Absence of gradient instability in DHOST scalar sector:\\$ \bFS>0 $};
	\node (DHOSTscalarSuperluminality) [requirements, right=2cm of Sigma] {Absence of ghost instability and superluminality	in DHOST scalar sector:\\$ \bGS\geq\bFS $};
	\node (ScaleInvariance) [requirements, right=2cm of P1] {Generation of nearly scale‐invariant entropy perturbations:\\$ n_s\simeq0.965 $};
	\node (Full) [requirements, right=2cm of chi] {Stability and non‐superluminality in full model and efficient conversion from entropy to curvature perturbations:\\$ \bFS>\bY P_1 \bg^2 $};
	\node (StabilityPotential) [requirements, right=2cm of P04] {Boundedness of the scalar potential from below};
	\node (EOM) [requirements, right=2cm of P02f012] {$ f_1+6f_2 $ and three background equations of motion};
	\graph [use existing nodes, edges={blue!70,line width=2pt}] {phi -> a -> a0g0 -> a1g1 -> Sigma -> P1 -> chi -> P04 -> P02f012};
	\graph [use existing nodes, edges={magenta!70,line width=3pt}] {a <- BKL};
	\graph [use existing nodes, edges={magenta!70,line width=3pt}] {a0g0 <- tensor};
	\graph [use existing nodes, edges={magenta!70,line width=3pt}] {a1g1 <- DHOSTscalarInstability};
	\graph [use existing nodes, edges={magenta!70,line width=3pt}] {Sigma <- DHOSTscalarSuperluminality};
	\graph [use existing nodes, edges={magenta!70,line width=3pt}] {P1 <- ScaleInvariance};
	\graph [use existing nodes, edges={magenta!70,line width=3pt}] {chi <- Full};
	\graph [use existing nodes, edges={magenta!70,line width=3pt}] {P04 <- StabilityPotential};
	\graph [use existing nodes, edges={magenta!70,line width=3pt}] {P02f012 <- EOM};
	\end{tikzpicture}
	\caption{Flowchart of the model construction using the inverse method. The left side shows the steps determining the Lagrangian functions, while the right side shows the corresponding constraints used. On the left side, the teal-colored rounded boxes are related the DHOST sector, while those in orange correspond to the extra scalar sector. \label{fig:flowchart}}
\end{figure}
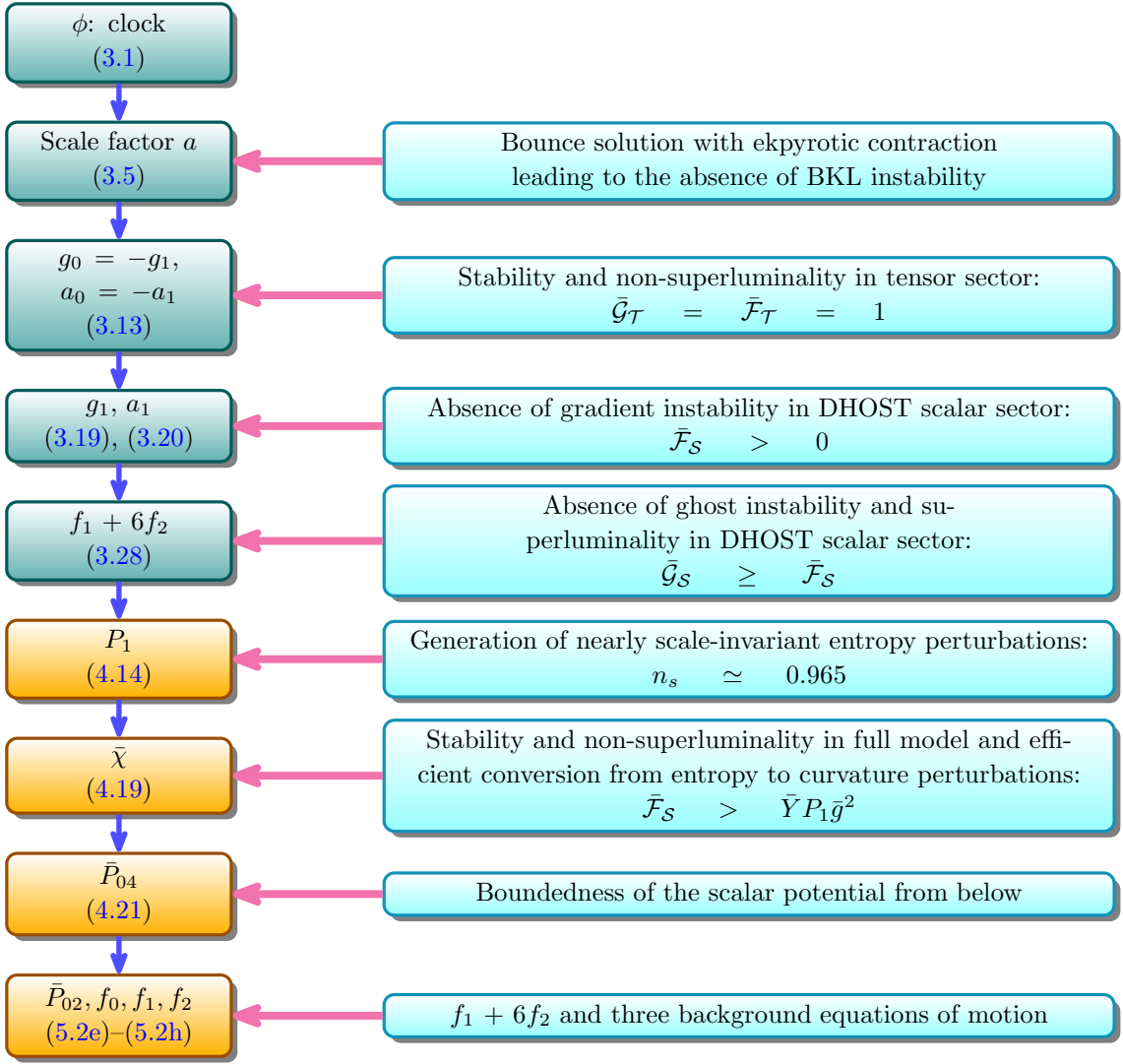
The result of the found Lagrangian functions in the action \eqref{eq:action} is given by
\begin{subequations}\label{eq:lagrangian-functions}
	\begin{align}
	&\mu^{-4}F(\phi,X)=\bF(\bphi,\bX) = f_0(\bphi)+f_1(\bphi)\bX+f_2(\bphi)\bX^2,\\
	&M_{pl}^{-2}F_2(\phi,X)=\bF_2(\bphi,\bX) = -\frac12+g_1(\bphi)-g_1(\bphi)\bX,\\
	&M_{pl}^{-2}\mu^4 A_1(\phi,X)=\bA_1(\bphi,\bX) = -a_1(\bphi)+a_1(\bphi)\bX,\\
	&\mu^{-1}K(\phi,X)=\bar K(\bphi, \bX) = 0,\\
	&\mu^{-4}P(\chi,Y,\phi)=\bP(\bchi,\bphi,\bY)=P_1(\bphi)\bY-\bP_{02}(\bphi)\bchi^2-\bP_{04}(\bphi)\bchi^4,
	\end{align}
\end{subequations}
where $ \bphi $, $ \bX $, $ \bchi $, $ \bY $, $ \bF $, $ \bF_2 $, $ \bA_1 $, $ \bP $, $ \bP_{02} $ and $ \bP_{04} $ are rescaled quantities (see \eqref{eq:rescaled-phi} and \eqref{eq:rescaled-chi}), and
\begin{subequations}\label{eq:lagrangian-functions-lower}
	\begin{align}
	& g_1(\bphi)=\frac{1}{3} w \cosh^{-2}\left(\frac{\bphi}{\tau }+u\right),\\
	& a_1(\bphi)=\frac{w}{12}\text{sech}^2\left(\frac{\bphi}{\tau }+u\right) \left(\frac{2 \left(\bphi^2 \tanh \left(\frac{\bphi}{\tau }+u\right)+\tau ^2 \tanh \left(\frac{\bphi}{\tau }\right)\right)}{\boldsymbol H(\bphi) \tau  \left(\bphi^2+\tau ^2\right)}-1\right),\\
	& P_1(\bphi) = \frac12\frac{\sigma \left(-\frac{\bphi}{\tau }\right)}{ \left(\frac{\bphi}{\tau }\right)^{2(1+b)}+1}+\frac{1}{2}\sigma \left(\frac{\bphi}{\tau }\right),\\
	& \bP_{04}(\bphi) = 10^{-10}\text{sech}\left(\frac{\bphi-\bt_c}{2\tau}\right),\\
	& \bP_{02}(\bphi)=-2\bP_{04}(\bphi)\boldsymbol\chi^2(\bphi)-\frac{1}{\boldsymbol a^3(\bphi) \boldsymbol\chi(\bphi)}\frac{d}{d\bphi}\left(\boldsymbol a^3(\bphi)\frac{d \boldsymbol\chi(\bphi)}{d\bphi} P_1(\bphi)\right),\label{eq:P02-explicit}\\
	& f_2(\bphi) = \frac{1}{8} \left(-30 a_1(\bphi) \frac{d g_1}{d\bphi} \boldsymbol H(\bphi)+2 a_1(\bphi) \frac{d \boldsymbol H}{d\bphi} g_1(\bphi)-6 a_1(\bphi) \frac{d \boldsymbol H}{d\bphi}+2 \frac{d a_1}{d\bphi} g_1(\bphi) \boldsymbol H(\bphi)\right.\nonumber\\
	&\hskip5em-6 \frac{d a_1}{d\bphi} \boldsymbol H(\bphi)+4 \frac{d^2 g_1}{d\bphi^2} g_1(\bphi)-6 \left(\frac{d g_1}{d\bphi}\right)^2+8 \frac{d g_1}{d\bphi} g_1(\bphi) \boldsymbol H(\bphi)-9 \frac{d g_1}{d\bphi} \boldsymbol H^2(\bphi) \theta(\bphi)\nonumber\\
	&\hskip5em +30 \frac{d g_1}{d\bphi} \boldsymbol H(\bphi)+2 \frac{d \boldsymbol H}{d\bphi} g_1(\bphi)^2-16 \frac{d \boldsymbol H}{d\bphi} g_1(\bphi)+2 \frac{d \boldsymbol H}{d\bphi}+9 g_1(\bphi) \boldsymbol H^2(\bphi)\nonumber\\
	&\hskip5em\left.+3 \boldsymbol H^2(\bphi)+2 P_1(\bphi)\left(\frac{d \boldsymbol\chi}{d\bphi}\right)^2 +2 \boldsymbol\Sigma(\bphi) \right),\label{eq:f2-explicit}\\
	& f_1(\bphi) = \frac{3}{2} h^2(\bphi) \left(3 \frac{d g_1}{d\bphi} \boldsymbol\Theta(\bphi) -3 g_1(\bphi)-1\right)-3 \frac{d \boldsymbol H(\bphi)}{d\bphi} g_1(\bphi)\nonumber\\
	&\hskip15em-\frac{d \boldsymbol H(\bphi)}{d\bphi}-2 f_2(\bphi)-P_1(\bphi) \left(\frac{d \boldsymbol\chi}{d\bphi}\right)^2,\label{eq:f1-explicit}\\
	& f_0(\bphi) = \bP_{02}(\bphi)\boldsymbol\chi^2(\bphi)+\bP_{04}(\bphi)\boldsymbol\chi^4(\bphi)-f_1(\bphi)\nonumber\\
	&\hskip10em-\left(\frac{d \boldsymbol\chi}{d\bphi}\right)^2 P_1(\bphi)-f_2(\bphi)-2 \frac{d h}{d\bphi}-3 h^2(\bphi),\label{eq:f0-explicit}
	\end{align}	
\end{subequations}
with
\begin{subequations}
	\begin{align}
	& \sigma(\bphi) = \frac{1}{1+\exp(-\bphi)},\\
	& \boldsymbol H(\bphi) = \frac{d}{d\bphi}\log \boldsymbol a(\bphi),\\
	& \boldsymbol a(\bphi)=\sigma \left(\frac{\bphi}{\tau }\right)\left[\left(\frac{\bphi}{\tau }\right)^2+1\right]^{\frac 16}+ \sigma \left(-\frac{\bphi}{\tau }\right) \left[\left(\frac{\bphi}{\tau }\right)^2+1\right]^{\frac{1}{2 \epsilon }},\\
	& \boldsymbol\Sigma(\bphi) = \frac{\tau ^4}{\left(\bphi^2+\tau ^2\right)^2}+\frac{(\epsilon -3) \left(1-\tanh \left(\frac{\bphi}{\tau }\right)\right)}{2 \epsilon ^2 \left(\bphi^2+\tau ^2\right)},\\
	& \boldsymbol\Theta(\bphi)=\boldsymbol H(\bphi) \left(4 a_1(\bphi)+g_1(\bphi)+1\right)+\frac{d g_1}{d\bphi},\\
	& \boldsymbol\chi(\bphi) = \bchi_0+\lambda\, \text{sech}\left(\frac{\bphi-\bt_c}{2 \tau }\right)\left[e^{-\left(\frac{\bphi-\bt_c}{4 \tau }\right)^2}-q\right].
	\end{align}
\end{subequations}
Here the parameter $ \tau $ controls the duration of the bounce phase, $ \epsilon $ characterizes the equation of state during the ekpyrotic contraction phase, $ b $ specifies the scalar spectral index, the characteristic energy scale $ \mu $ determines the amplitude of the curvature perturbations, and $ \bt_c $ characterizes the center of conversion period. The parameters $ u $ and $ w $ have been introduced to avoid the fine-tuning problem that $ H $ and $ \Theta $ cross zero at the same time. The parameters $ q $, $ \bchi_0 $ and $ \lambda $ are related to the efficiency of the conversion process. We numerically checked that there are no instabilities and superluminality for the parameters given by \eqref{eq:parameters-full} and the characteristic energy scale $ \mu\sim\cO(10^{-2})M_{pl} $, see e.g. Figs. \ref{fig:gs-fs}, \ref{fig:cssquared} and \ref{fig:positivity-fab}. 
	
In summary the model constructed in this work gives rise to a completely healthy and phenomenologically viable bounce scenario, which satisfies the followings:
\begin{itemize}
	\item it is free of BKL, ghost and gradient instabilities,
	\item it has conventional asymptotics for DHOST sector,
	\item it is free of superluminality,
	\item it predicts nearly scale-invariant curvature perturbations, generated by an entropic mechanism, compatible with observations,
	\item it predicts a sufficiently low tensor-to-scalar ratio.
\end{itemize}

From the phenomenological point of view, it is important to make sure that the non-Gaussianities of the curvature perturbations do not exceed the observational bounds. It is actually obvious that the perturbations of the entropic field $ \chi $ have negligible non-Gaussianities \cite{Fertig:2013yyy,Ijjas:2014fja}. Thus, the main contribution to the non-Gaussianities of the curvature perturbations should come from the non-linear conversion process from the entropic to curvature perturbations. Since our scenario has efficient (almost $ 100\% $ of efficiency) and smooth conversion from entropy to curvature perturbations, we expect that the non-Gaussianities would not be generated too much \cite{Fertig:2013yyy,Fertig:2015ola,Fertig:2016czu}. In any case it would be interesting to confirm this point. We plan to pursue this direction in future work.

\newpage

\appendix

\section{Background equations of motion}\label{app:background}

This Appendix gives the background equations of motion for spatially-flat FLRW metric, which are obtained by varying the action \eqref{eq:action} with respect to $ \alpha $, $ \zeta $ and $ \chi $, respectively:
\begin{itemize}
	\item $ \alpha $:
	\begin{align}\label{eq:background-g00}
	& F+P-2YP_Y-6 F_2 H^2-6 H \dot\phi \left(F_{2\phi}-\left(A_1-2 F_{2X}\right) \ddot\phi\right)\nonumber\\
	&+\dot\phi^2 \left(-6 A_1 \dot H+A_3 \ddot\phi^2+A_4 \ddot\phi^2-2 F_X+12 F_{2X} \dot H+24 F_{2X} H^2+K_{\phi }\right)\nonumber\\
	&+\dot\phi^3 \left(H \left(-6 A_{1\phi}+3 (A_3+2 A_4) \ddot\phi-6 K_X\right)+2 (A_3+A_4) \dddot\phi\right)\nonumber\\
	& +\dot\phi^4 \left(3 H^2 \left(4 A_{1X}+3 A_3\right)+\ddot\phi \left(\ddot\phi \left(2 A_{3X}+2 A_{4X}+3 A_5\right)+2 A_{3\phi}+2 A_{4\phi}\right)+3 A_3 \dot H\right)\nonumber\\
	&+\dot\phi^5 \left(3 H \left(A_{3\phi}+2 A_5 \ddot\phi\right)+2 A_5 \dddot\phi\right)+2 \dot\phi^6 \ddot\phi \left(A_{5X} \ddot\phi+A_{5\phi}\right)=0
	\end{align}
	\item $ \zeta $:
	\begin{align}\label{eq:background-gii}
	& 2 A_1 \ddot\phi^2+F+P-4 F_2 \dot H-6 F_2 H^2-4 F_{2X} \ddot\phi^2-2 F_{2\phi} \ddot\phi\nonumber\\
	&+2 \dot\phi \left[\dddot\phi \left(A_1-2 F_{2X}\right)-2 H \left(2 \left(F_{2X}-A_1\right) \ddot\phi+F_{2\phi}\right)\right]\nonumber\\
	&- \dot\phi^2 \left[-4 A_1 \dot H-6 A_1 H^2-4 A_{1X} \ddot\phi^2-2 A_{1\phi} \ddot\phi+2 A_3 \ddot\phi^2-A_4 \ddot\phi^2\right.\nonumber\\
	&\hskip5em\left.+8 F_{2XX} \ddot\phi^2+8 F_{2\phi X} \ddot\phi+2 F_{2\phi\phi}+2 K_X \ddot\phi+K_{\phi } \right]\nonumber\\
	&+\dot\phi^3 \left[4 H \left(2 A_{1X} \ddot\phi+A_{1\phi}\right)-A_3 \dddot\phi\right]+\dot\phi^4 \ddot\phi \left[\left(A_5-2 A_{3X}\right) \ddot\phi-A_{3\phi}\right]=0
	\end{align}
	\item $ \chi $:
	\begin{align}\label{eq:background-chi}
	\frac{1}{a^3}\frac{d}{dt}\left(a^3\dot\chi P_Y\right)-\frac12 P_\chi=0.
	\end{align}
\end{itemize}

One can readily check that in the case of single-field Beyond Horndeski theories where $ P=0 $,
$ A_1=-X F_4-2 G_{4X} $, $ A_3=-2F_4 $, $ A_4=2F_4 $ and $ A_5=0 $ the above background equations of motion coincide with those given in \cite{Mironov:2018oec}.

\section{Exceptional Subclass}\label{app:exceptional-case}

For the exceptional sublcass of DHOST Ia theories, the Lagrangian functions $ A_4 $ and $ A_5 $ become
\begin{subequations}\label{eq:Exceptional-A4-A5}
	\begin{align}
	A_4= &\;-\frac{2 \left(A_1-2 F_{2X}\right) \left(9 A_1^2 X^2-17A_1 F_2 X+2 X F_{2X} (7
		F_2-6A_1 X)+8 F_2^2\right)}{X (3A_1 X-4 F_2)^2},\\
	A_5 = &\; \frac{2 \left(A_1-2 F_{2X}\right){}^2}{X (3A_1 X-4 F_2)}.
	\end{align}
\end{subequations}
Inserting the ansatz \eqref{eq:ansatz} into \eqref{eq:Exceptional-A3} and \eqref{eq:Exceptional-A4-A5}, we obtain
\begin{subequations}\label{eq:A3-A4-A5}
	\begin{align}
	A_3=&\; \frac{2 \left[a_0(\phi )+X a_1(\phi )+2 g_1(\phi )\right] \left[X (a_0(\phi )+X a_1(\phi )+2 g_1(\phi ))+2 g_0(\phi )+1\right]}{X \left[X (3 a_0(\phi )+3 X a_1(\phi )+4 g_1(\phi ))+4 g_0(\phi )+2\right]},\\
	A_4 = &\; -\frac{(a_0(\phi )+X a_1(\phi )+2 g_1(\phi )) }{X (X (3 a_0(\phi )+3 X a_1(\phi )+4 g_1(\phi ))+4 g_0(\phi )+2)^2}\times\nonumber\\
	&\hskip3em\times\left[X a_0(\phi ) \left(36 X^2 a_1(\phi )+34 g_0(\phi )+58 X g_1(\phi )+17\right)+18 X^2 a_0(\phi )^2\right.\nonumber\\
	&\hskip4em+18 X^4 a_1(\phi )^2+X^2 a_1(\phi ) (34 g_0(\phi )+58 X g_1(\phi )+17)\nonumber\\
	&\hskip4em\left.+2 (2 g_0(\phi )+2 X g_1(\phi )+1) (4 g_0(\phi )+11 X g_1(\phi )+2)\right],\\
	A_5 = &\; \frac{2 \left[a_0(\phi )+X a_1(\phi )+2 g_1(\phi )\right]^2}{X \left[X (3 a_0(\phi )+3 X a_1(\phi )+4 g_1(\phi ))+4 g_0(\phi )+2\right]}.
	\end{align}
\end{subequations}

\bibliographystyle{JHEP3}
\bibliography{DHOST_bounce_stability}

\end{document}